\documentclass[twocolumn,trackchanges]{aastex7}

\usepackage[titletoc]{appendix} 
\usepackage{graphicx}
\usepackage{dcolumn}
\usepackage{bm}
\usepackage{amsmath}
\usepackage{overpic}
\usepackage{booktabs}%
\usepackage{makecell}
\usepackage{enumerate}
\usepackage[figuresright]{rotating}
\makeatletter

\newcommand{\Rmnum}[1]{\expandafter\@slowromancap\romannumeral #1@}
\makeatother
\usepackage{hyperref}
\usepackage{xcolor}
\usepackage{longtable}


\begin{document}

\title{GRB 240715A: Revealing Novel Intrinsic Mechanism by Different Individual Pulse} 

\correspondingauthor{Shao-Lin Xiong}
\email{xiongsl@ihep.ac.cn}

\author[]{Wen-Jun Tan}
\affil{State Key Laboratory of Particle Astrophysics, Institute of High Energy Physics, Chinese Academy of Sciences, Beijing 100049, China}
\affil{University of Chinese Academy of Sciences, Chinese Academy of Sciences, Beijing 100049, China}
\email{}

\author{Chen-Wei Wang}
\affil{State Key Laboratory of Particle Astrophysics, Institute of High Energy Physics, Chinese Academy of Sciences, Beijing 100049, China}
\affil{University of Chinese Academy of Sciences, Chinese Academy of Sciences, Beijing 100049, China}
\email{}

\author{Shao-Lin Xiong*}
\affil{State Key Laboratory of Particle Astrophysics, Institute of High Energy Physics, Chinese Academy of Sciences, Beijing 100049, China}
\email{}

\author{Shi-Jie Zheng}
\affil{State Key Laboratory of Particle Astrophysics, Institute of High Energy Physics, Chinese Academy of Sciences, Beijing 100049, China}
\email{}

\author{Jiang He}
\affil{State Key Laboratory of Particle Astrophysics, Institute of High Energy Physics, Chinese Academy of Sciences, Beijing 100049, China}
\affil{University of Chinese Academy of Sciences, Chinese Academy of Sciences, Beijing 100049, China}
\email{}

\author{Xiao-Yun Zhao}
\affil{State Key Laboratory of Particle Astrophysics, Institute of High Energy Physics, Chinese Academy of Sciences, Beijing 100049, China}
\email{}

\author{Yue Huang}
\affil{State Key Laboratory of Particle Astrophysics, Institute of High Energy Physics, Chinese Academy of Sciences, Beijing 100049, China}
\email{}

\author{Shu-Xu Yi}
\affil{State Key Laboratory of Particle Astrophysics, Institute of High Energy Physics, Chinese Academy of Sciences, Beijing 100049, China}
\email{}

\author{Bing Li}
\affil{State Key Laboratory of Particle Astrophysics, Institute of High Energy Physics, Chinese Academy of Sciences, Beijing 100049, China}
\email{}

\author{He Gao}
\affil{Institute for Frontier in Astronomy and Astrophysics, Beijing Normal University, Beijing 102206, China, Department of Astronomy, Beijing Normal University, Beijing 100875, China}
\affil{Department of Astronomy, Beijing Normal University, Beijing 100875, China}
\email{}

\author{Bo-bing Wu}
\affil{State Key Laboratory of Particle Astrophysics, Institute of High Energy Physics, Chinese Academy of Sciences, Beijing 100049, China}
\email{}

\author{Bing Zhang}
\affil{The Hong Kong Institute for Astronomy and Astrophysics, University of Hong Kong, Fokfulam Road, Hong Kong, China}
\affil{Department of Physics, the University of Hong Kong, Pokfulem Road, Hong Kong, China}
\affil{Nevada Center for Astrophysics, University of Nevada, Las Vegas, NV 89154, USA}
\email{}

\author{Fr\'ed\'eric Daigne}
\affil{Sorbonne Universit\'e, CNRS, UMR 7095, Institut d’Astrophysique de Paris, 98 bis bd Arago, F-75014 Paris, France}
\email{}

\author{Maria-Grazia Bernardini}
\affil{INAF—Osservatorio Astronomico di Brera, Via Emilio Bianchi 46, 23807 Merate, Italy}
\email{}

\author{Bin-Bin Zhang}
\affil{School of Astronomy and Space Science, Nanjing University, Nanjing 210093, China}
\affil{State Key Laboratory of Modern Astronomy and Astrophysics (Nanjing University), Ministry of Education, China}
\affil{Purple Mountain Observatory, Chinese Academy of Sciences, Nanjing 210023, China}
\email{}

\author{Stéphane Basa}
\affil{Aix Marseille Univ, CNRS, CNES, LAM, Marseille, France}
\email{}

\author{Bertrand Cordier}
\affil{CEA Paris-Saclay, Irfu/D\'epartement d'Astrophysique, 91191 Gif sur Yvette, France}
\email{}

\author{Jin-Song Deng}
\affil{School of Astronomy and Space Science, University of Chinese Academy of Sciences, Beijing 101408, China}
\affil{National Astronomical Observatories, Chinese Academy of Sciences, Beijing 100101, China}
\email{}

\author{Yong-Wei Dong}
\affil{State Key Laboratory of Particle Astrophysics, Institute of High Energy Physics, Chinese Academy of Sciences, Beijing 100049, China}
\email{}

\author{Damien Dornic}
\affil{CPPM, Aix Marseille Univ, CNRS/IN2P3, CPPM, Marseille, France}
\email{}

\author{Olivier GODET}
\affil{Institut de Recherche en Astrophysique $\&$ Plan\'etologie, Universit\'e de Toulouse/CNRS/CNES, 9 avenue du colonel Roche, 31028 Toulouse, France}
\email{}

\author{Xu-Hui Han}
\affil{National Astronomical Observatories, Chinese Academy of Sciences, Beijing 100101, China}
\email{}

\author{Mao-Hai Huang}
\affil{Chinese Academy of Sciences, Beijing 100049, China}
\email{}

\author{Cyril Lachaud}
\affil{APC, Univ.Paris Diderot, CNRS/IN2P3, CEA/lrfu, Obs de Paris, Sorbonne Paris Cit\'e, France}
\email{}

\author{Hua-Li Li}
\affil{National Astronomical Observatories, Chinese Academy of Sciences, Beijing 100101, China}
\email{}

\author{Pierre Maggi}
\affil{Observatoire Astronomique de Strasbourg, Université de Strasbourg, CNRS, 11 rue de l’Universit\'e, F-67000 Strasbourg, France}
\email{}

\author{Fr\'ed\'eric Piron}
\affil{LUPM, CC 72, CNRS/IN2P3, Universite de Montpellier, place E. Bataillon 34095 Montpellier CEDEX 05, France}
\email{}

\author{Yu-Lei Qiu}
\affil{National Astronomical Observatories, Chinese Academy of Sciences, Beijing 100101, China}
\email{}

\author{Stéphane Schanne}
\affil{CEA Paris-Saclay, Irfu/D\'epartement d'Astrophysique, 91191 Gif sur Yvette, France}
\email{}

\author{Li-Ming Song}
\affil{State Key Laboratory of Particle Astrophysics, Institute of High Energy Physics, Chinese Academy of Sciences, Beijing 100049, China}
\email{}

\author{Damien Turpin}
\affil{CEA Paris-Saclay, Irfu/D\'epartement d'Astrophysique, 91191 Gif sur Yvette, France}
\email{}

\author{Susanna Vergani}
\affil{LUX, Observatoire de Paris, PSL University, CNRS, Sorbonne University, 92190 Meudon, France}
\email{}

\author{Jing Wang}
\affil{National Astronomical Observatories, Chinese Academy of Sciences, Beijing 100101, China}
\email{}

\author{Ping Wang}
\affil{State Key Laboratory of Particle Astrophysics, Institute of High Energy Physics, Chinese Academy of Sciences, Beijing 100049, China}
\email{}

\author{Jian-Yan Wei}
\affil{National Astronomical Observatories, Chinese Academy of Sciences, Beijing 100101, China}
\affil{University of Chinese Academy of Sciences, Chinese Academy of Sciences, Beijing 100049, China}
\email{}

\author{Chao Wu}
\affil{National Astronomical Observatories, Chinese Academy of Sciences, Beijing 100101, China}
\email{}

\author{Li-Ping Xin}
\affil{National Astronomical Observatories, Chinese Academy of Sciences, Beijing 100101, China}
\email{}

\author{Yang Xu}
\affil{National Astronomical Observatories, Chinese Academy of Sciences, Beijing 100101, China}
\email{}

\author{Li Zhang}
\affil{State Key Laboratory of Particle Astrophysics, Institute of High Energy Physics, Chinese Academy of Sciences, Beijing 100049, China}
\email{}

\author{Shuang-Nan Zhang}
\affil{State Key Laboratory of Particle Astrophysics, Institute of High Energy Physics, Chinese Academy of Sciences, Beijing 100049, China}
\affil{University of Chinese Academy of Sciences, Chinese Academy of Sciences, Beijing 100049, China}
\email{}

\begin{abstract}

The Space-based multiband astronomical Variable Objects Monitor (\textit{SVOM}), detected its first short gamma-ray burst, GRB 240715A, in-flight, which was jointly observed by \textit{Fermi}. Based on observational data of \textit{SVOM}/GRM and \textit{Fermi}/GBM, we perform a comprehensive temporal and spectral analysis for individual pulse in the prompt emission of this burst, and novel characteristics are revealed. Firstly, opposite evolutions of spectral lag are found in the first and third pulse of this burst. Second, the large negative lag of the first pulse is an outlier in short GRB sample, especially when the pulse duration is considered. Spectral analysis shows that the negative lag of the first pulse is caused by the evolution of spectrum index, and is irrelevant to $E_{peak}$, which is inconsistent with the previous study. The intrinsic mechanism is probably attributed to electron cooling in the decaying magnetic field, which leads to the continuous hardening of the spectrum index and results in negative lag. Furthermore, spectral analysis also shows that the third pulse is more likely to be described by a quasi-thermal spectrum, indicating the existence of photospheric emission. It is difficult to explain how the synchrotron radiation appears before photospheric emission in a single GRB and some assumptions are discussed.

\end{abstract}


\section{Introduction}


\textit{SVOM} (Space-based multiband astronomical Variable Objects Monitor) is a Sino-French mission committed to Gamma-Ray Burst (GRB) studies, and to the discovery and multi-wavelength follow-up of cosmic transients \citep{SVOM_intro, SVOMdescribe}. The mission, which was launched on June 22nd 2024, includes a satellite with a multi-wavelength science payload, and a dedicated ground segment made of robotic telescopes in China and Mexico. The satellite carries two wide-field monitors for GRB prompt emission detection, called ECLAIRs (a 4-150 keV coded-mask imager, \citep{ECL_intro}) and GRM (a 15-5000 keV Gamma-Ray Monitor, \citep{GRM_intro}). ECLAIRs is in charge of providing precise localization ($<$ 13 arcmin at 90$\%$ confidence level) and X-ray energy cover, while GRM is in response of obtaining the broad energy spectrum measurement. 
Besides these two monitors, two narrow-field telescopes for GRB follow-up are also equipped, called MXT (Microchannel X-ray Telescope, energy between 0.2 and 10 keV, \citep{MXT_intro}) and VT (Visible Telescope, from 450 to 650 nm for the blue channel and  from 650 to 1000 nm for the red channel, \citep{SVOM_intro}).

Gamma-ray bursts (GRBs) are the most violent explosions in the universe. Traditionally, GRBs can be divided into long GRBs and short GRBs by a simple but effective criterion---$T_{90}$ \citep{twoclassGRB, T90classification}. As research deepens, more physical classifications have been proposed, corresponding to two different origin models for GRBs, namely Type I (usually short) and Type II (usually long) GRBs \citep{Zhang2006type}. Generally speaking, Type I bursts are produced by the merger of compact binaries with shorter durations and harder spectra \citep{Blinnikovmerger1984,1986Paczynskiph,1Meszarosmerger1992}, while Type II bursts are related to the core collapse of supermassive stars, which are usually longer and softer. \citep{Woosleycollapsar1993,Paczycollapsar1998,Woosleycollapsar2006}.
    
GRB 240715A is the first short GRB that triggered the GRM in-flight \citep{GRM_trigger} with real-time burst message processing. The prompt emission of this burst reveals novel characteristics. Firstly, it exhibits an opposite evolution of the spectral lag in two pulses, which means there is a relatively large negative spectral lag excluding statistical fluctuations in the first pulse. Furthermore, this GRB also shows a possible photospheric emisssion in latter pulse. Both phenomena require new mechanism explanation that cannot be covered by current models.

The spectral lag of gamma-ray bursts (GRBs), attributed to the fact that higher-energy gamma-ray photons arrive earlier than lower-energy gamma-ray photons, is a quite common characteristic in prompt emission of GRBs \citep{1995chenglag,1996NorrisLag}. Some studies showed that long GRBs usually have larger positive spectral lags, whereas short GRBs tend to exhibit negligible, sometimes negative, lags \citep{2000NorrisLag,2006Yilag}. 


Some papers argued that negative spectral lags may be unreal, because if there are two nearly overlapping pulses with the second one being harder than the first one, then fake negative lags will appear when calculating the lag of the whole light curve \citep{2014INTEGRALlag,2011penglag}. Therefore, lag calculation for pulses in GRBs has also been studied. \cite{2018LuLag} mainly studied pulses of long GRBs and gave the correlation of observed quantities, such as lag-energy, lag-pulse width and lag index-pulse width. \cite{2006Yilag} systematically studied the pulse lag of short GRBs and found that a proportion of $\sim$ 17\% of short GRBs has a negative spectral lag excluding the statistical fluctuation effect, which is significantly different from that of long GRBs.

Several models are proposed to interpret the spectral lag. These models fall broadly into two categories. One invokes the kinematic process, the so-called ``high-latitude” effect (the curvature effect), which expresses that photons at higher latitudes would arrive at the observer later with softened observed energy due to a smaller Doppler factor, resulting in a lag. \cite{2005shenlag,2006lulag} studied this effect in detail 
and found that the curvature effect marginally interpreted the observed lags. Other models mainly considered the dynamic process, which expresses that the spectral evolution during the prompt phase of the GRB can explain the observed lag \citep{2005rydelag,2003Kocevskilag,2011penglag,2000wulag,2018LuLag}. Because of the continuous cooling effect of radiated electrons, the peak energy ($E_{peak}$) of the energy spectra moves from high to low energies in gamma-rays, resulting in the peak of the light curve also moving from higher energy band to the lower one, further resulting in the observed spectral lag. 
Based on a simple but general emission model, \cite{2016Uhmlag} interpreted some observed spectral lag features by considering the intrinsic curved spectral shape, the decay of the magnetic field strength, and the rapid bulk acceleration of the emission zone. The above models aimed to explain the positive lag. However, a small fraction of the observed negative lags is still in debate, the relatively popular model is inverse Comptonization \citep{2014Roychoudhurylag, 2024Vyaslag}.

There are two leading scenarios that explain the radiation mechanism that powers the prompt emission of GRBs. The first is  the synchrotron radiation model, which is relevant for either the internal shock model \citep{1990ShemiIS,1992ReesIS,1993MeszarosIS,1994ReesIS,2000Lloydsyn,2004Baringsyn}, or magnetic dissipation driven by Poynting flux  \citep{1992Usovpoy,1994Thompsonpoy,1997meszpoy,2003Lyutikovpoy,2003Vlahakispoy}, or internal-collision-induced magnetic reconnection and turbulence (ICMART) at a large radius \citep{2011Zhangicmart,2015Dengicmart}. The second scenario is known as the photospheric emission model produced from relativistic outflows, as a natural consequence of the fireball model \citep{1986Goodmanph,1986Paczynskiph}. In this model, the optical depth at the base of the outflow is significantly greater than unity \citep{1999PiranPh}, and a bright quasi-thermal spectral component is expected. 

In the context of the photospheric emission model, the standard blackbody spectrum (Planck spectrum) is broadened for two main reasons. First, other dissipative processes such as synchrotron radiation or Comptonization of thermal photons occur beneath the photosphere, modifying the shape of blackbody spectrum \citep{2005APeerph,2006APeerph,2005Reesph}. Second, the geometric broadening of the structured jet can also modify the shape of blackbody spectrum \citep{1991Abramowiczph,2008Peerph}. In this latter scenario, the luminosity, Lorentz factor, and photosphere radius are found to be a function of the angle .\citep{2013Lundmanph,2018Mengph,2019Mengph,2021Wangph}. Consequently, the observed spectrum is a superposition of comoving blackbodies with different temperatures, originating from various angles relative to the line of sight \citep{2010Rydeph,Hou2018mbb}. This structural broadening results in a predominance of low-energy photons and produces a softer spectrum, typically described as a multicolor blackbody. 
Furthermore, a hybrid jet model was proposed to interpret cases with superposition of a dominant synchrotron component and a sub-dominant thermal component \citep{2012Axelsson,2013Guiriecph,2015Gaoph,2022Mengph,2025wang0825}. In addition, transition from fireball to Poynting-flux-dominated outflow has also been observed \citep{2018Zhang160625}.

This paper is organized as follows. Obsevations and data reduction are described in Section \ref{section2}. Detailed temporal and spectrum analysis is presented in Section \ref{section3}. A brief discussion and explanation of the observed phenomena of this GRB is presented in Section \ref{section4}. Summary is given in Section \ref{section5}. 
    
Throughout the paper, a standard cosmology with parameters $H_0$= 67.4 km \,$\rm{s^{-1}\,Mpc^{-1}}$, $\Omega_M$=0.315, and $\Omega_\Lambda$= 0.685 is adopted \citep{cosmology}. All parameter errors in this work are for the 68\% confidence level if not otherwise stated.
    
\section{Observations and Data Reduction}\label{section2}

As a scintillator-based instrument, GRM consists of three detectors (GRD Gamma-Ray Detectors) operating in the 15 keV--5 MeV energy range. The three identical detectors are inclined by 30 degrees with respect to the axis of the ECLAIRs telescope and spaced from each other in the perpendicular plane of 120 degrees \citep{SVOMdescribe,GRM_intro}.


GRB 240715A triggered \textit{SVOM}/GRM at 05:44:03.400 UT on 15 July 2024 in-flight with all three GRDs \citep{GCN0715A}, when SVOM was in the commissioning phase. It is the first short GRB that triggered the GRM in-flight \citep{GRM_trigger}. At that time, ECLAIRs was not in operation. The duration ($T_{90}$) in the 15--5000 keV is about 0.17 \,s. We extract 15--5000 keV light curve with a time-bin of 5 ms from event-by-event data combined from all 3 GRDs of GRM. The light curve shows mainly three pulses, especially in the time-energy diagram (Fig.~\ref{fig:lc}a). Moreover, we extract the light curve in 7 energy bands: 15--30 keV, 30--70 keV, 70--100 keV, 100--300 keV, 300--500 keV, 500--2000 keV and 2000--5000 keV, and multi-wavelength light curves are shown in Fig.~\ref{fig:lc}b.
For GRM spectrum analysis, we extract the energy spectrum of all three detectors of GRM using GECAMTools-v20240514.
\footnote{\href{https://github.com/zhangpeng-sci/GECAMTools-Public}
{https://github.com/zhangpeng-sci/GECAMTools-Public}}
The energy range used for spectral fitting is 20--5000 keV for GRM detectors. Hereafter, we take the ground trigger time (2024-07-15T05:44:03.000) as T$_{0}$.

GRB 240715A also triggered \textit{Fermi}/GBM \citep{GCN0715AGBM}. The Gamma-ray Burst Monitor (GBM) is one of the two instruments onboard the \textit{Fermi} Gamma-ray Space Telescope \citep{GBM_meegan_09,GBM_Bissaldi_09}, which consists of 14 detectors with different orientations: 12 Sodium Iodide (NaI) scintillation detectors (labeled from n0 to nb) and 2 Bismuth Germanate (BGO) scintillation detectors (b0 and b1). For GBM spectrum analysis, we extract the energy spectrum of NaI detectors with incident angle smaller than 40$^\circ$ (that is, n9 and na) and the BGO detector b1 using GBM data tools \footnote{\href{https://fermi.gsfc.nasa.gov/ssc/data/analysis/gbm/gbm_data_tools/gdt-docs/}
{https://fermi.gsfc.nasa.gov/ssc/data/analysis/gbm/gbm\_data\_tools/gdt-docs/}} \citep{2025ApJwang}.
The energy range used for spectra fitting is 8--900\,keV for NaI detectors and 0.3--35\,MeV for BGO detectors. 

\textit{Swift}/XRT conducted a taget of Opportunity (ToO) for GRB 240715A about 33 minutes after T${_0}$, but only a likely afterglow has been detected and some upper limits are given \citep{GCN0715AXRT}. Some optical follow-up observations were also conducted. However, no new optical source was detected consistent within the XRT position and a few upper limits are given \citep{GCN0715AOP1,GCN0715AOP2}. 
The absence of an optical counterpart and of the identification of the possible host galaxy prevented the redshift determination.

\section{data analysis}\label{section3}
The first pulse and the third pulse of this GRB exhibit different behavior in temporal and spectral domains, including negative spectral lag in the first pulse and possible photospheric emission in the third pulse.

\subsection{Pulse fitting}\label{pulse_fitting}

\begin{figure*}
\centering
\begin{tabular}{cc}
\begin{overpic}[width=0.37\linewidth]{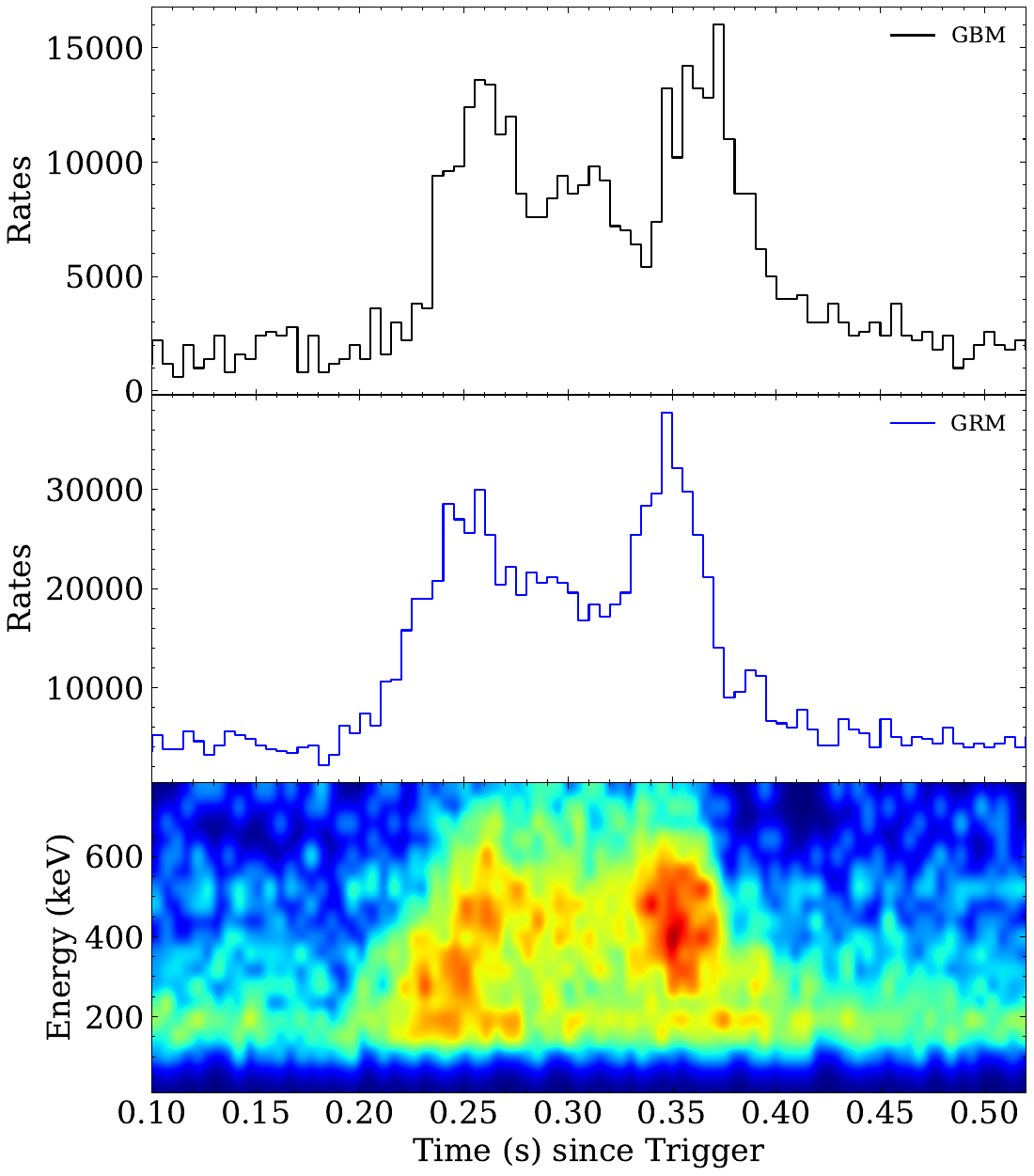}\put(0, 98){\bf a}\end{overpic} 
\begin{overpic}[width=0.47\linewidth]{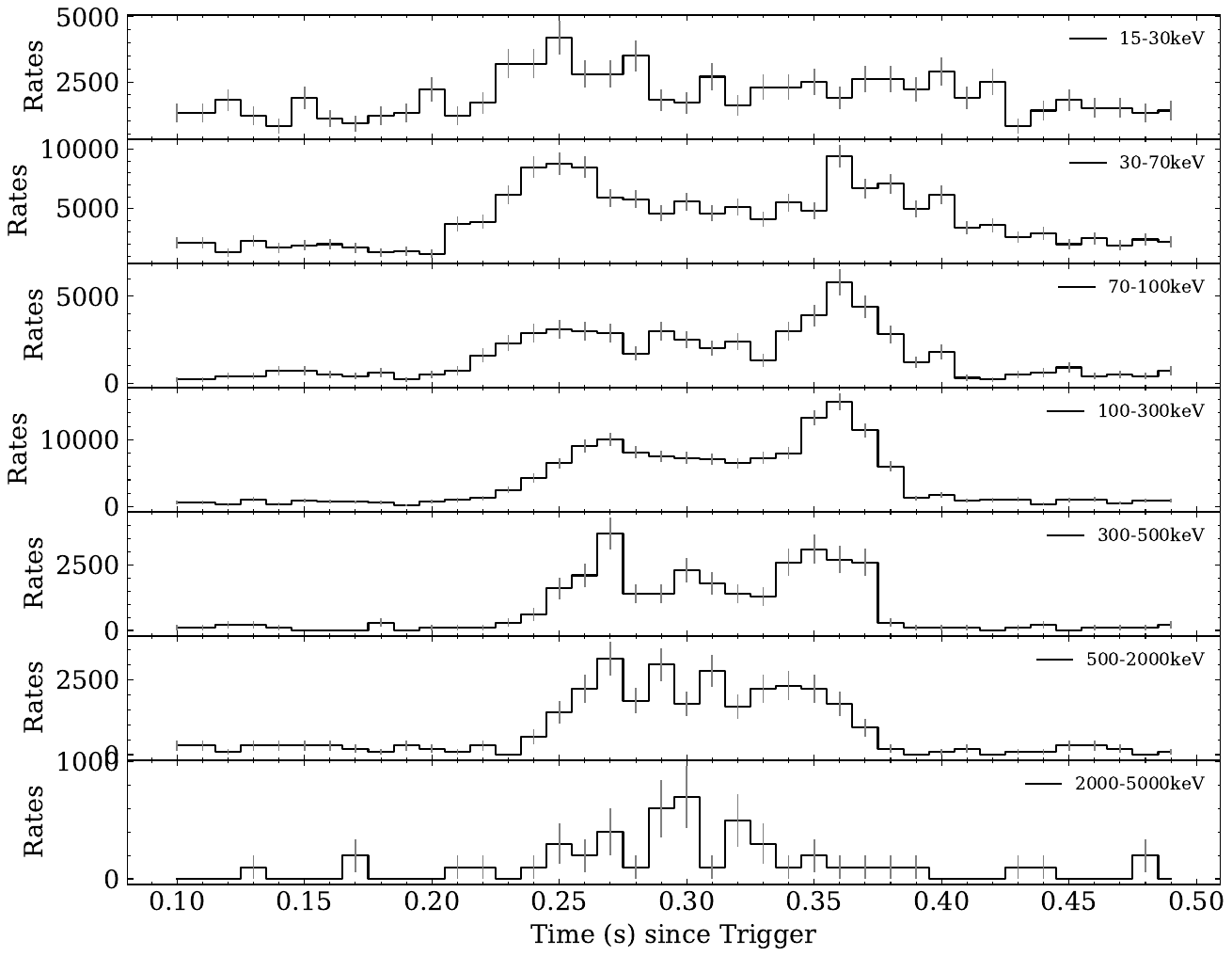}\put(0, 95){\bf b}\end{overpic} \\
\\
\begin{minipage}[b]{0.48\linewidth}
    \begin{overpic}[width=\textwidth]{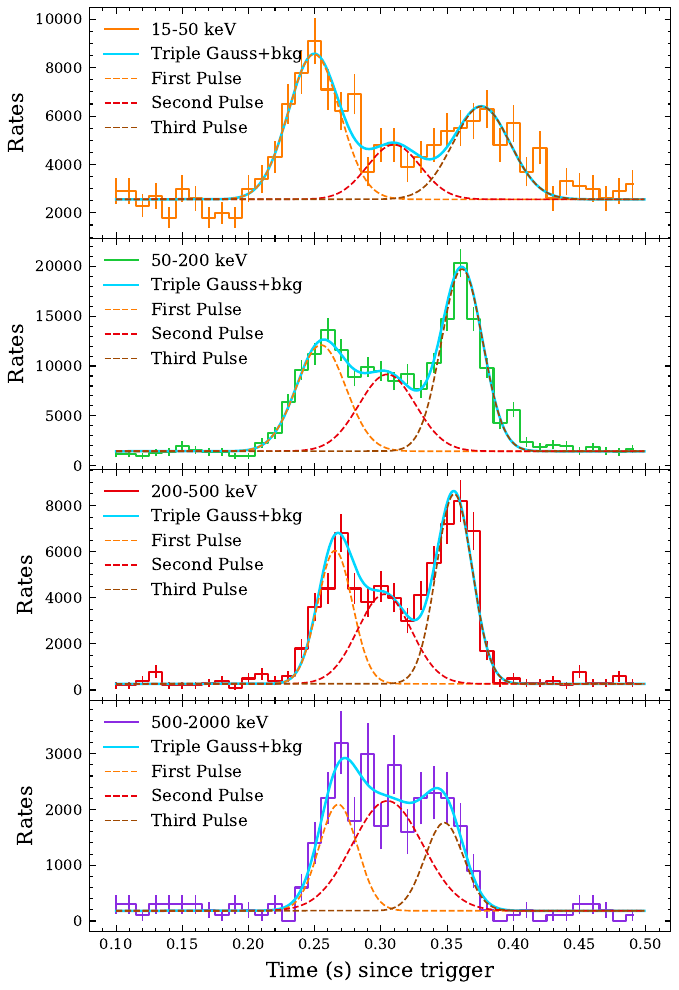}\put(0, 97){\bf c}\end{overpic}
\end{minipage}
\begin{minipage}[b]{0.36\linewidth}
        \begin{overpic}[width=\textwidth]{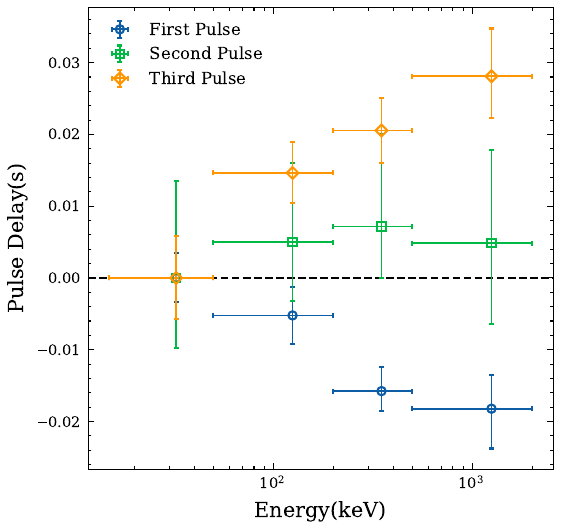}\put(0, 93){\bf d}\end{overpic} 
        \begin{overpic}[width=\textwidth]{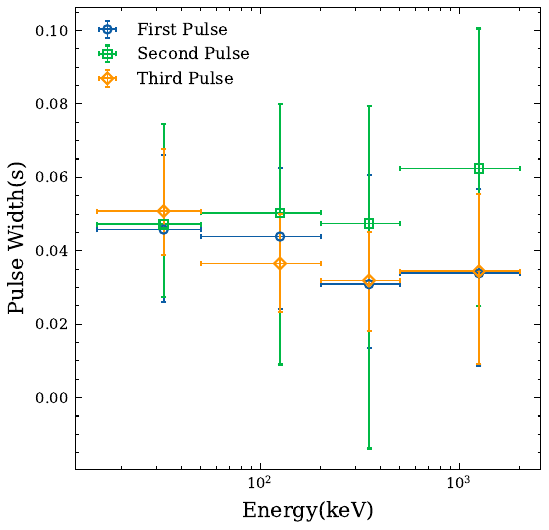}\put(0, 93){\bf e}\end{overpic}
\end{minipage}

\end{tabular}
\caption{\noindent\textbf{Light curves and pulse fitting. }
\textbf{a}, Light curves of \textit{SVOM}/GRM and \textit{FERMI}/GBM with 5 ms bin width, and time-energy diagram of \textit{SVOM}/GRM within the energy range from 10 keV to 1000 keV.  \textbf{b}, Multi-wavelength light curves of \textit{SVOM}/GRM with 10 ms bin width. \textbf{c}, The triple-pulse fitting of the light curves in each engergy band with triple-gaussian function.  \textbf{d}, The time delay of the peak of each pulse.  \textbf{e}, The width of each pulse in each energy band .}
\label{fig:lc}
\end{figure*}

In order to study the pulse properties, it is very important to separate different pulses. From the light curve in Fig.~\ref{fig:lc}a, GRB 240715A has two bright well-shaped pulses, which lie at the left and right ends of the light curve, respectively. This paper focuses mainly on these two pulses. As the shape of these two pulses is relatively narrow and symmetrical, we choose two Gaussian functions to fit these two pulses, respectively. Meanwhile, it can also be seen from the light curve that there is a small pulse-like structure between these two pulses. Therefore, we used another Gaussian function to fit this small pulse in the middle. We used a triple-Gaussian to fit the three pulses of the light curve:
\begin{equation}   
L=N_{1}G(t,t_{p1},s_{1})+N_{2}G(t,t_{p2},s_{2})+
N_{3}G(t,t_{p3},s_{3})+B
\label{equ:pules_fit}
\end{equation}

Where $G(t,t_{p},s) = {\rm exp}(-(t-t_{p})^2/s)$. $N_{1}$, $N_{2}$, $N_{3}$, $t_{p1}$, $t_{p2}$ and $t_{p3}$ are the normalizations, peak times of three pulses. $s = 2\sigma^2$, where $\sigma$ is the standard deviation of Gaussion function. B is the constant background. The reason why constant background is chosen is that we first fitted the whole light curve using first-order polynomial and found that the normalization of the first-order term was very small (from 10$^{-3}$ to 10$^{-2}$). Moreover, the time range used for pulse fitting is very short (T${_0}$-0.1 s to T${_0}$+0.5 s), so there was no significant changes in the background. 

We extract the light curves in four energy bands, i.e., 15--50 keV, 50--200 keV, 200--500 keV and 500--2000 keV with a time-bin of 10 ms from event-by-event data combined from all three GRDs of GRM, and then fit the light curve in each energy band with triple-Gaussian function, (Fig.~\ref{fig:lc}c). The fitting parameters are listed in Table~\ref{tab:Pulse_fit_pars}. 
Then we calculate the time delay of the peak in the lowest energy band relative to that in each high energy band for each Gaussian model (Fig.~\ref{fig:lc}d)
and the pulse width of each energy band (Fig.~\ref{fig:lc}e), which is defined as the full width at half maxima (FWHM) of the Gaussian function. As the energy increases, the first pulse peaks at later times, whereas the third pulse peaks at lower times, and the peak of the middle pulse does not move basically. Additionally, as the energy increases, the widths of the first pulse and the third pulse are slightly reduced, but the width of the middle pulse is basically unchanged. 
Combined with pulse-peak shifting and pulse width reducing, the whole picture looks like that the first pulse and the third pulse slowly approach to each other as the energy increases, and finally enhance the brightness of the second pulse in the high energy band.

\subsection{Spectral lag}

Based on the pulse fitting, we can determine at which time period a certain pulse is dominant, and then determine the start and end times of each pulse for calculating the spectral lag (mainly the end time of the first pulse and start time of the third pulse). We calculate the spectral lag of the first pulse (T$_{0}$+0.1 s to T$_{0}$ to 0.28 s), the second pulse (T$_{0}$+0.28 s to T$_{0}$+0.33 s), the third pulse (T$_{0}$+0.33 s to T$_{0}$+0.5 s) and the whole light curve (T$_{0}$+0.1 s to T$_{0}$+0.5 s), respectively. To investigate in detail the pulse behavior of each energy band, we estimate the spectral lag with respect to 15–-30 keV for six energy bands based on the signal significance: 30--70 keV, 70--100 keV, 100--300 keV, 300--500 keV and 500--2000 keV of all three GRDs of GRM. Considering the signal significance, the light curve above 2000 keV is not choosen.

The modified cross-correlation function (CCF), more suitable for transient signals like GRBs, proposed by \cite{1997BANDCCF} is adopted to compute the spectral lag:

\begin{equation}   
CCF(k\Delta t;c_{1},c_{2}) = \frac{ {\textstyle \sum_{i=max(1,1-k)}^{min(N,N-k)}}c_{1i}c_{2(i+k)}}{\sqrt{ {\textstyle \sum_{i}^{}}\,c_{1i}^2 {\textstyle \sum_{i}^{}}\,c_{2i}^2}}    ,
\label{equ:Bandccf}
\end{equation}

where $\Delta t$ is the time bin of light curve, $k\Delta t$ ($k$ = .., -1, 0, 1, ...) is a multiple of the time bin and represents the time delay, c1 and c2 are the count rates of two light curve in each energy band. We use this formula to calculate the CCF for a series of time delays $k\Delta t$ and determine the spectral lag as the time delay that corresponds to the maximum of the CCF versus time delay \citep[see e.g.,][]{Bernardinilag2015}. We take two small time intervals before and after the burst to perform background subtraction, as the background did not change much during the burst. We adopt 1 ms as the temporal binning to calculate the CCF, as the burst is bright enough. To estimate the uncertainty on the spectral lag, we generate 1000 Monte Carlo (MC) simulated light curves based on the observed initial light curve with Poisson probability distribution and calculate the lag of each simulated light curve \citep{2010Ukwattalag}. Then we take the 1$\sigma$ uncertainty of the distribution as the uncertainty on the spectral lag. 

\begin{figure*}[htbp]
\begin{minipage}[b]{0.45\linewidth}
    \centering
    \begin{tabular}{c}
        \begin{overpic}[width=\textwidth]{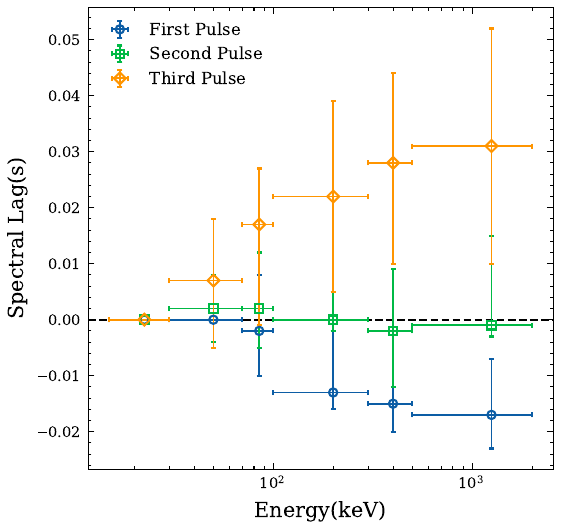}\put(-2, 90){\bf a}\end{overpic}
        \begin{overpic}[width=\textwidth]{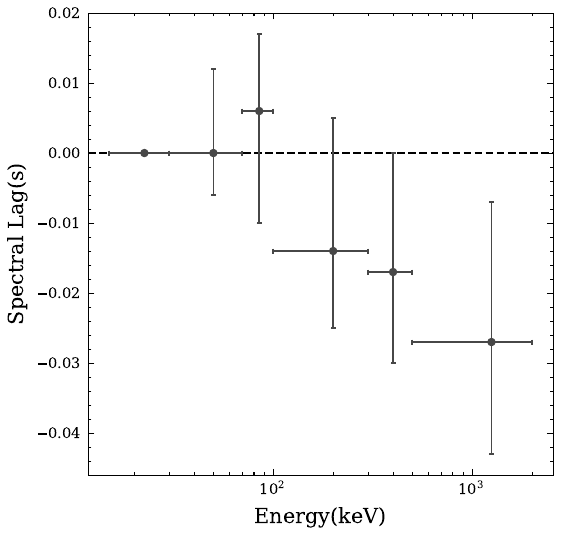}\put(0, 90){\bf b}\end{overpic}
    \end{tabular}
\end{minipage}\\
\\
\begin{minipage}[b]{0.56\linewidth}
    \centering
    \begin{tabular}{c}
        \begin{overpic}[width=\textwidth]{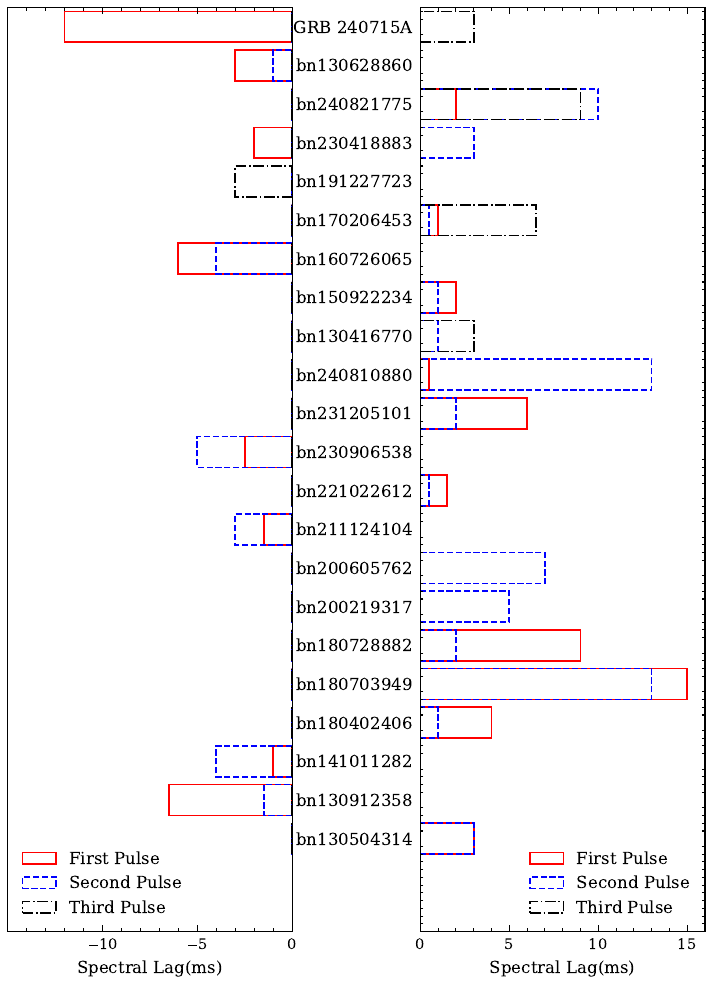}\put(-2, 98){\bf c}\end{overpic}
    \end{tabular}
\end{minipage}
\begin{minipage}[b]{0.34\linewidth}
    \centering
    \begin{tabular}{c}
        \begin{overpic}[width=\textwidth]{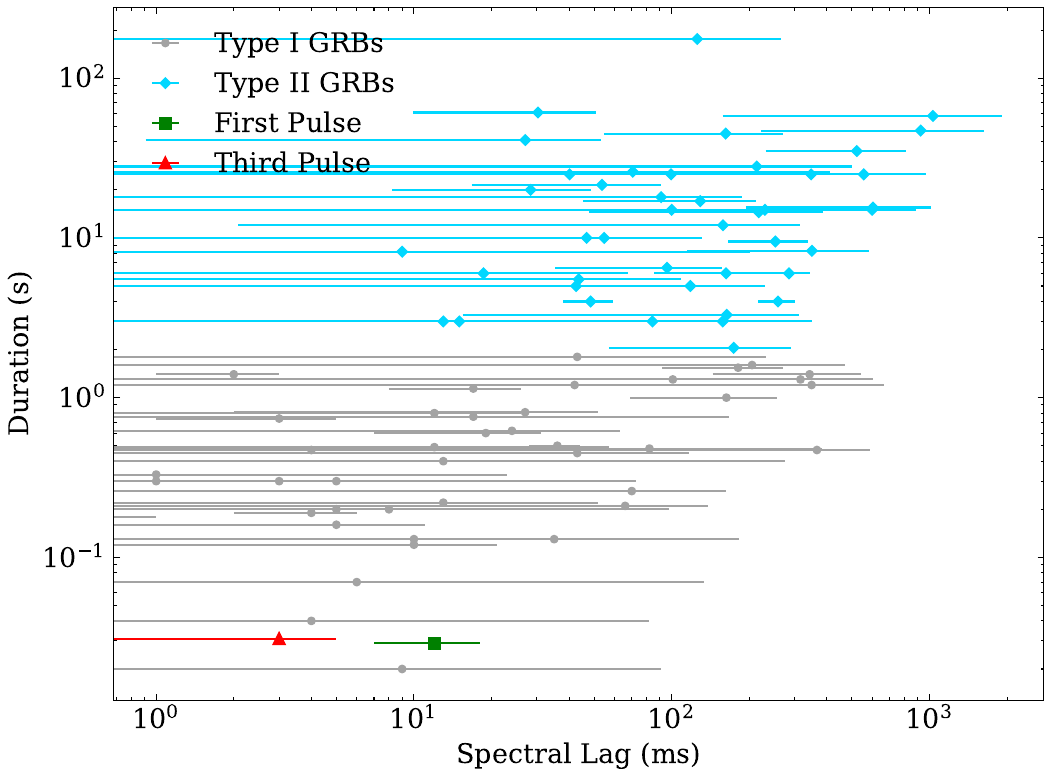}\put(0, 76){\bf d}\end{overpic} \\
        \begin{overpic}[width=\textwidth]{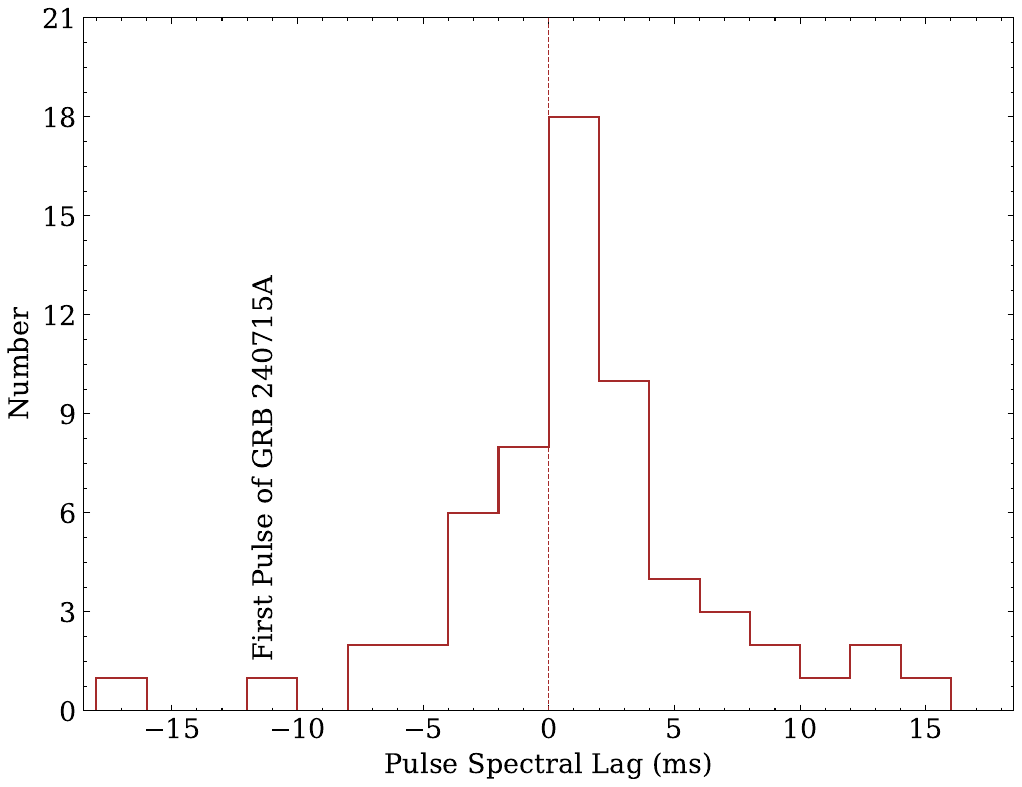}\put(0, 76){\bf e}\end{overpic}\\
        \begin{overpic}[width=\textwidth]{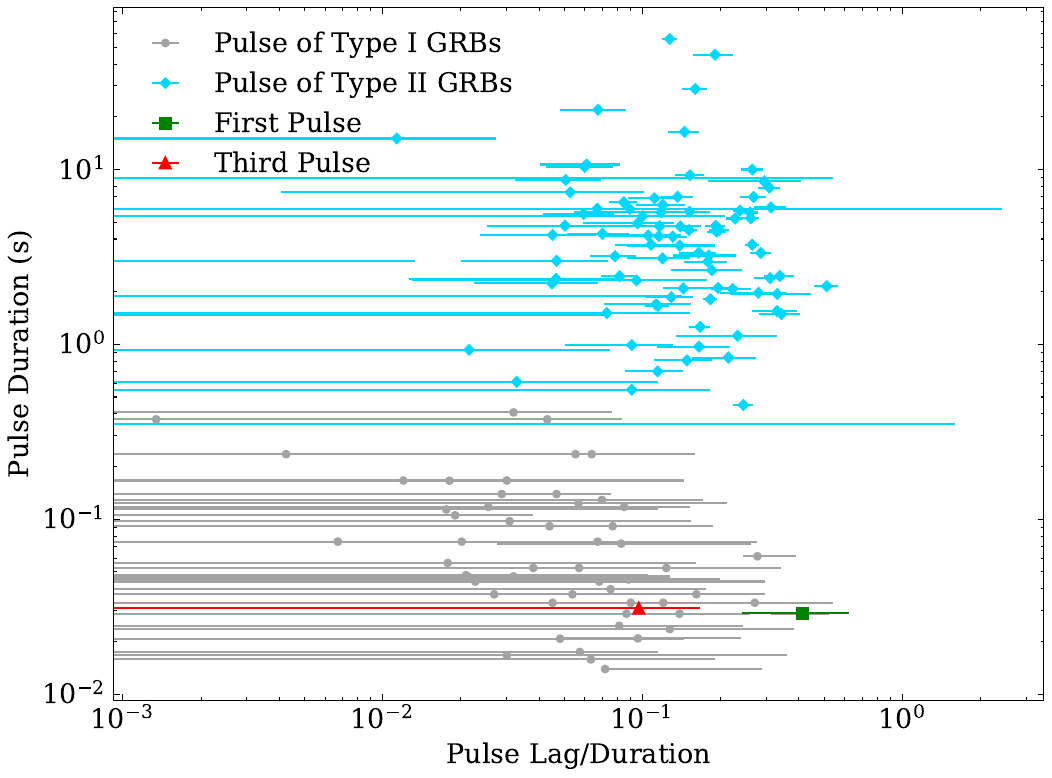}\put(0, 76){\bf f}\end{overpic}\\
    \end{tabular}
\end{minipage}
\caption{\textbf{a}, Spectral lag of the each pulse of GRB 240715A. \textbf{b}, Spectral lag of the whole light curve of GRB 240715A. \textbf{c}, Spectral lags of each pulse for short GRBs from the GBM sample. The bars in different colors represent the lag of different pulses. Left panel shows the negative lags of pulses in GRBs, right panel shows the positive lags of pulses in GRBs.  \textbf{d}, The burst duration versus burst lag (in absolute value). The GRB sample is from \cite{Bernardinilag2015,GehrelsLAG2006,Goldstein0817,Xiaolag2022}.  \textbf{e}, The histogram of lags of pulse in short GRBs, and the lag of the first pulse of GRB 240715A is labeled. \textbf{f}, The pulse duration versus the ratio of pulse lag and pulse duration (in absolute value). The Type II GRB sample is from \cite{2018LuLag}}
\label{fig:lag_of_each_pulse}
\end{figure*}

The result shows that the first pulse has a relatively large negative lag, the middle pulse almost exhibits no spectral lag, and the third pulse has a normal positive lag (Fig.~\ref{fig:lag_of_each_pulse}a), which means that in the first pulse low energy photons arrive earlier than high energy photons, in the second pulse low energy photons and high energy photons arrive simultaneously, and the third pulse has the opposite behavior with respect to the first pulse. Moreover, the negative lag of the first pulse increases with energy, as well as the positive lag of the third pulse, forming a phenomenon that the lag of both pulses evolve with energy symmetrically (Fig. \ref{fig:lag_of_each_pulse}a). Finally, the opposite evolution of the spectral lag in the first pulse and the third pulse together results in an inverted ``V" shape of the overall spectral lag shape of the whole light curve (Fig.~\ref{fig:lag_of_each_pulse}b). 

From the above analysis, the spectral lag of GRB 240715A has two characteristics: one is that two pulses with opposite spectral lag evolution occur in the same GRB; the other is that the negative lag of the first pulse is relatively large, especially considering the duration (which is expressed as the width of the pulse in our work) of the pulse. In order to further study the rarity and uniqueness of this behavior, as well as to avoid the fact that the brightness of different pulses varying at different energy bands would cause the overall lag to be misestimated \citep{2014ITEGRALcat}, we search the bright short bursts which consist of well-shaped pulse in the \textit{Fermi GBM Burst Catalog} \footnote{\href{https://heasarc.gsfc.nasa.gov/W3Browse/fermi/fermigbrst.html}{https://heasarc.gsfc.nasa.gov/W3Browse/fermi/fermigbrst.html}}, just like some work that focus on long bursts \citep{2018LuLag}. These short bursts are either composed of single-pulse, double-pulse or triple-pulse. We fit each pulse of these GRBs with Gaussian function and calculate the width of these pulses in the 50--300 keV band, as well as the spectral lag of these pulses between energy band of 50--100 keV and 150--300 keV. The width and lag results are listed in Table~\ref{tab:lag_width}. As shown in Fig. \ref{fig:lag_of_each_pulse}c and Table~\ref{tab:lag_width}, most bright short bursts with well-shaped pulses do not have evident spectral lags within the errors, but a small fraction of pulses of short bursts indeed exhibit a negative lag excluding statistical fluctuations, just as stated in \cite{2006Yilag}. None of the short bursts considered exhibit the same spectral lag behaviour as seen in GRB 240715A, except for GRB 240715A. Another exception is bn230418883, however, the lag is so small that it can be regarded as no opposite evolution when errors are taken into account ($-2^{+3}_{-1}$\,ms and $+3^{+6}_{-5}$\,ms for each pulse). It is clearly shown that it is very rare for the spectral lag of two pulses in the same GRB to evolve in opposite. 

In addition, as shown in Fig.~\ref{fig:lag_of_each_pulse}e, the negative lag of the first pulse in GRB 240715A ($-12^{+5}_{-6}$\,ms) is relatively large among the short GRB sample, which is about twice as long as the normal negative lag in other GRBs. 
Such a relatively large negative lag is even more peculiar when the duration of the pulse is considered. The pulse lag is not large in the sample of the whole GRB light curve, as shown in Fig.~\ref{fig:lag_of_each_pulse}d. However, when both the lag and duration of the pulse are considered, it is clearly that the ratio of lag (absolute value) to the duration of pulse of GRB 240715A is among the highest in our sample, as well as the sample in \cite{2006Yilag}, especially the negative lag of the first pulse. Even though there are several pulses with negative lags in other GRBs, it is truly unique to have such a large negative lag in such a short duration. As shown in Fig.~\ref{fig:lag_of_each_pulse}f, the ratio of lag (absolute value) to the duration of the first pulse in GRB 240715A is close to 0.5, means that half of the apparent duration of the first pulse is dominated by the delay of the high energy emission relative to the low energy emission. Furthermore, it can be seen from Fig. ~\ref{fig:lag_of_each_pulse}f that in the pulse of short bursts, the shorter the duration, the larger the ratio tends to be.

\subsection{Spectral analysis} \label{section3.3}

\begin{figure*}[htbp]
\centering
\begin{minipage}[b]{0.6\linewidth}
    \centering
    \begin{tabular}{c}
        \begin{overpic}[width=\textwidth]{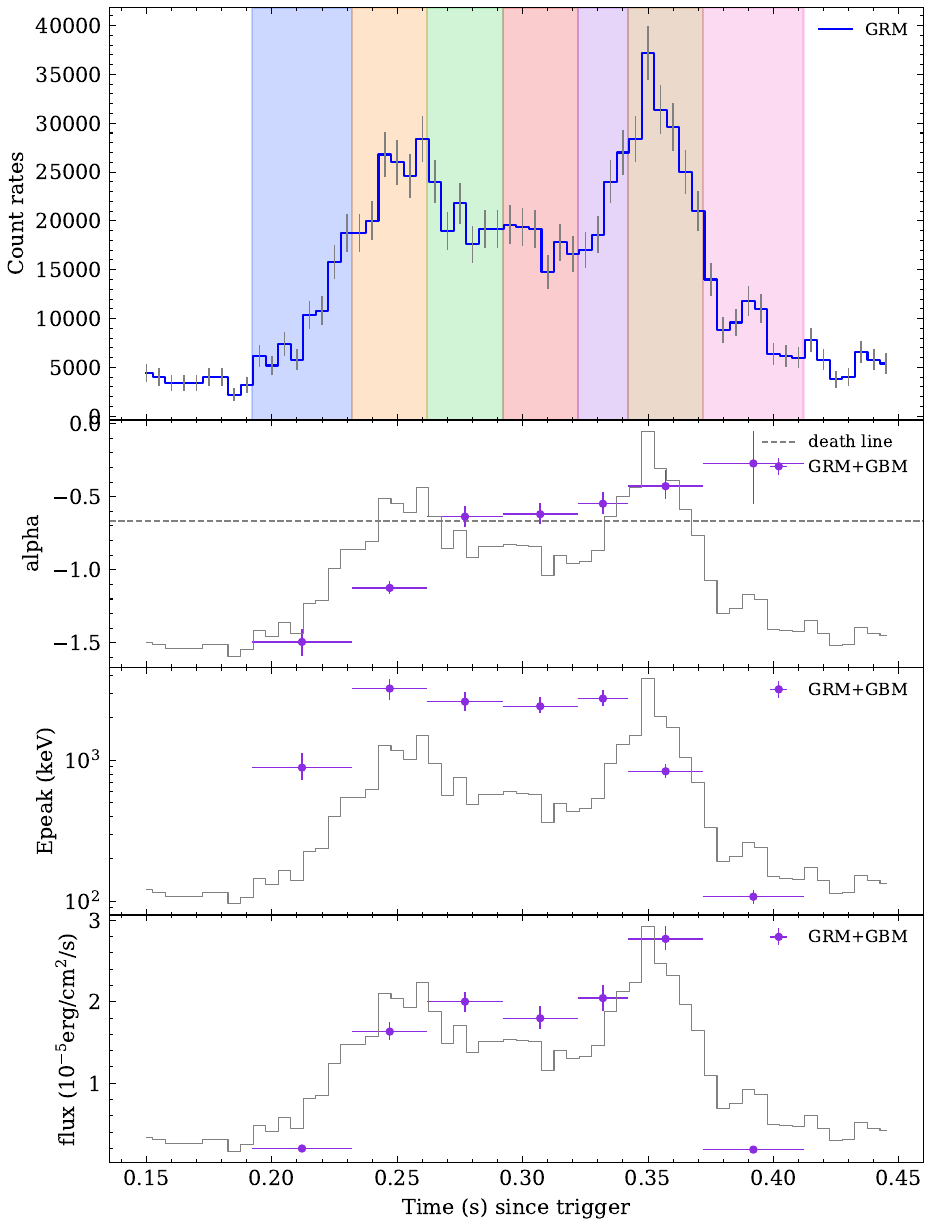}\put(0, 98){\bf a}\end{overpic}
    \end{tabular}
\end{minipage}
\begin{minipage}[b]{0.35\linewidth}
    \centering
    \begin{tabular}{c}
        \begin{overpic}[width=\textwidth]{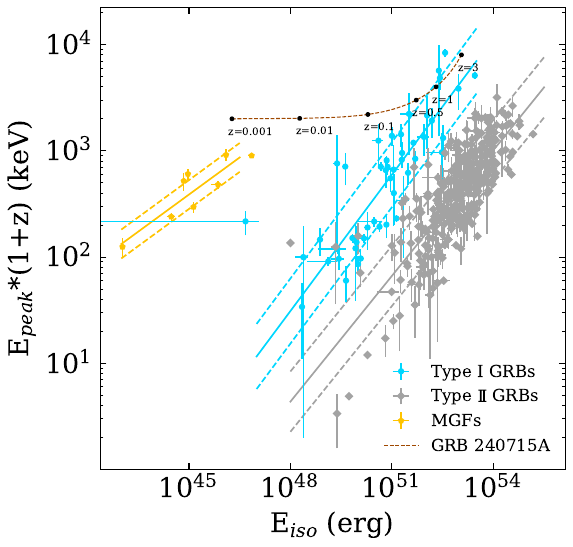}\put(0, 90){\bf b}\end{overpic} \\
        \\
        \\
        \\
        \\
        \begin{overpic}[width=\textwidth]{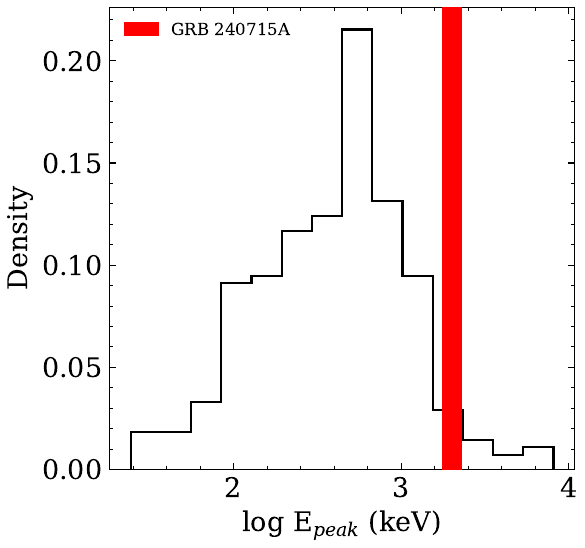}\put(0, 90){\bf c}\end{overpic}
    \end{tabular}
\end{minipage}
\caption{{\bf Spectrum fitting resluts.}
\textbf{a}, Top panel: Light curve of GRM and the seven slices for the fine time-resolved spectrum were marked with different colors. The following three panels are respectively the photon index, Epeak and flux of each slice fitted with GRM plus GBM data. \textbf{b}, The $E_{p,z}$ and $E_{\rm iso}$ correlation diagram. The best-fit for Type II (gray points), Type I (blue points) GRBs and MGFs (yellow points) are plotted (solid lines) with the 1$\sigma$ boundary (dashed line) marked. The red dashed line represent the evolution of the $E_{p,z}$ and $E_{\rm iso}$ of GRB 240715A as the redshift changes from 0.001 to 3. The GRB sample is from \cite{Lan23}, and the MGF sample is from  \cite{2020zhangMGF,2024MereghettiMGF} \textbf{c}, E$_{peak}$ distribution of short GRBs observed by \textit{Fermi}/GBM and GRB 240715A. The red bar represents $E_{peak}$ and error of GRB 240715A. The $E_{peak}$ data reference to \cite{2017luEp}. }
\label{fig:spec}
\end{figure*}

We use the Pyxspec software \citep{pyxspec} to conduct the spectral analysis. The time-integrated spectrum of the GRB 240715A prompt emission is computed from $T_{0}$+0.18 s to $T_{0}$+0.43 s, and is well fitted by a cutoff power-law (CPL) model with a power-law photon index $\alpha = -0.86\pm 0.04$, and $E_{peak} = 2000_{-231}^{+273}$ keV. The $E_{peak}$ value found for GRB 240715A is on the higher end of the  $E_{peak}$ distribution built from \textit{Fermi}/GBM short GRBs, but still not an outlier (Fig.~\ref{fig:spec}c). Moreover, since there is no redshift measurement for GRB 240715A, we plot the rest-frame peak energy $E_{p,z} -E_{iso}$ curve at different redshifts ranging from 0.001 to 3 in the Amati relation \citep{Amati2002} diagram (Fig.~\ref{fig:spec}b). It is worth noting that apart from the traditional Type I and Type II GRBs, a third track was recently found for magnetar giant flares (MGFs) \citep{2020YangMGF,2021SvinkinMGF,2021RobertsMGF,2020zhangMGF,2024MereghettiMGF}. We find that GRB 240715A is consistent with Type I bursts with redshift in the range of 0.1 to 3.

For the time-resolved spectrum, we first divide the light curve into three coarse time slices based on the pulse profile, which are denoted as S-I. In order to investigate the evolution of energy spectrum in detail, we further divide the light curve into seven fine time slices (denoted as S-II). The boundaries of each slice of S-I and S-II are listed in Table~\ref{tab:spectral_of_GRB}.

For each slice, we perform a detailed spectral fit  with the CPL model, the ratio of Profile-Gaussian likelihood to the degree of freedom (PGSTAT/dof, \cite{1996pgstat}) is taken into account to test the goodness-of-fit. CPL model is expressed as:

\begin{equation}   
N(E)=A\left(\frac{E}{E_0}\right)^{\alpha} {\rm exp}(-\frac{E}{E_{\rm c}}),
\label{equ:CPL_Model}
\end{equation}
where $A$ is the normalization constant ($\rm photons \cdot cm^{-2} \cdot s^{-1} \cdot keV^{-1}$), $\alpha$ is the power law photon index, $E_0$ is the pivot energy fixed at 1 keV, and $E_{\rm c}$ is the characteristic cutoff energy in keV. The peak energy $E_{\rm p}$ is related to $E_{\rm c}$ through $E_{\rm p}$=$(2 + \alpha)E_{\rm c}$.

The best-fit parameters obtained by the CPL models are listed in Table~\ref{tab:spectral_of_GRB}, and the corresponding spectral evolution of fine time slices is plotted in Fig.~\ref{fig:spec}. The evolution of the energy spectrum parameters is relatively strong, and two interesting phenomena are displayed. First, the photon index exhibits a very regular evolution from soft to hard over time, and systematically exceeds the synchrotron ``death line”  \citep{1998syndeathline} defined by $\alpha$ = --2/3 in the time slice of the third pulse, reaching the highest value of $\sim-0.27$. Second, previous studies of time-resolved spectra of simple pulses in GRBs indicate that peak energy displays two prevailing trends: (1) hard-to-soft tracking, which means peak energy monotonically decays independently of the flux evolution; (2) intensity-tracking, meaning peak energy generally follows the trajectory of the flux \citep{1995ApJFordEp,2010ApJluEp,2012ApJluEp}. However, in this GRB, $E_{peak}$ remains basically unchanged after reaching its maximum value in the first pulse and exhibits a ``hard to soft" spectral evolution pattern in the third pulse.

\subsection{Possible photospheric emission of the third pulse} 

Since the photon index exceeds the ``death line" in the third pulse, we study the spectra of the third pulse (0.33--0.39 s) in detail. In addition to the CPL model, we also fit the data with the BAND model, the BAND-Cut model, the combination of CPL and Blackbody (expressed as bbodyrad) model, and a quasi-thermal spectral model, namely, the multicolor black body model (mBB, \cite{Hou2018mbb}). 

The BAND model is expressed as \citep{Band1993ApJ}:
\begin{equation}
N(E)=\left\{
\begin{array}{l}
A(\frac{E}{100\,{\rm keV}})^{\alpha}{\rm exp}(-\frac{E}{E_{\rm c}}),\,E<(\alpha-\beta)E_{\rm c}, \\
A\big[\frac{(\alpha-\beta)E_{\rm c}}{100\,{\rm keV}}\big]^{\alpha-\beta}{\rm exp}(\beta-\alpha)\\\,\,\,\,\,(\frac{E}{100\,{\rm keV}})^{\beta}, E\geq(\alpha-\beta)E_{\rm c}, 
\end{array}\right.
\label{equ:band_Model}
\end{equation}
where $A$ is the normalization constant ($\rm photons \cdot cm^{-2} \cdot s^{-1} \cdot keV^{-1}$), $\alpha$ and $\beta$ are the low-energy and high-energy power law spectral indices, $E_{\rm c}$ is the characteristic energy in keV and the peak energy $E_{\rm p}$ is related to the $E_{\rm c}$ through $E_{\rm p}$=$(2 + \alpha)E_{\rm c}$.

The BAND-Cut model is expressed as:
\begin{equation}
N(E)=\left\{
\begin{array}{l}
A{E}^{\alpha_{\rm 1}}{\rm exp}(-\frac{E}{E_{\rm 1}}),\,E<E_{\rm b}, \\
A{E_{b}}^{\alpha_{\rm 1}-\alpha_{\rm 2}}{\rm exp}(\alpha_{\rm 2}-\alpha_{\rm 1}){E}^{\alpha_{\rm 2}}{\rm exp}(-\frac{E}{E_{\rm 2}}), E\geq E_{\rm b}, 
\end{array}\right.
\label{equ:bandcut_Model}
\end{equation}
where $A$ is the normalization constant ($\rm photons \cdot cm^{-2} \cdot s^{-1} \cdot keV^{-1}$), $\alpha_{\rm 1}$ and $\alpha_{\rm 2}$ represent the spectral indices of the low-energy power law segment and the medium-energy power law segment smoothly connected at the break energy $E_{\rm b} = \frac{E_{\rm 1}E_{\rm 2}}{E_{\rm 2}-E_{\rm 1}}(\alpha_{\rm 1}-\alpha_{\rm 2})$. The peak energy of the $\nu F\nu$ spectrum is defined as $E_{\rm p}$=$(2 + \alpha_{\rm 2})E_{\rm 2}$, which locates at the high-energy exponential cutoff.

The bbodyrad(BB) model is expressed as
\begin{equation}   
N(E)=\frac{1.0344 \times 10^{-3}\times AE^2}{{\rm exp}(\frac{E}{kT})-1}
\label{equ:bb_Model}
\end{equation}
where $A=R^2_{km}/D^2_{10}$ is the normalization constant, $R_{km}$ is the source radius in km and $D_{10}$ is the distance to the source in units of 10 kpc, and $kT$ is the temperature in keV.

The multicolor black body (mBB) model is expressed as

\begin{equation}
    N(E)=\frac{8.0525(m+1)K}{[(\frac{T_{\rm max}}{T_{\rm min}})^{m+1}-1]}\left ( \frac{kT_{\rm min}}{\rm keV} \right )^{-2} I(E)
\label{equ:mbb_Model}
\end{equation}
where
\begin{equation}
    I(E)=(\frac{E}{kT_{\rm min}})^{m-1}\int_\frac{E}{kT_{\rm max}}^\frac{E}{kT_{\rm min}}\frac{x^{2-m}}{e^{x}-1}dx
\end{equation}

where $x=E/kT$, $m$ is the power-law index of the dependence of the luminosity distribution on the temperature, and the temperature ranges from the minimum $T_{\rm min}$ to the maximum $T_{\rm max}$.

\begin{figure*}
\centering
\begin{tabular}{cc}
\begin{overpic}[width=0.45\textwidth]{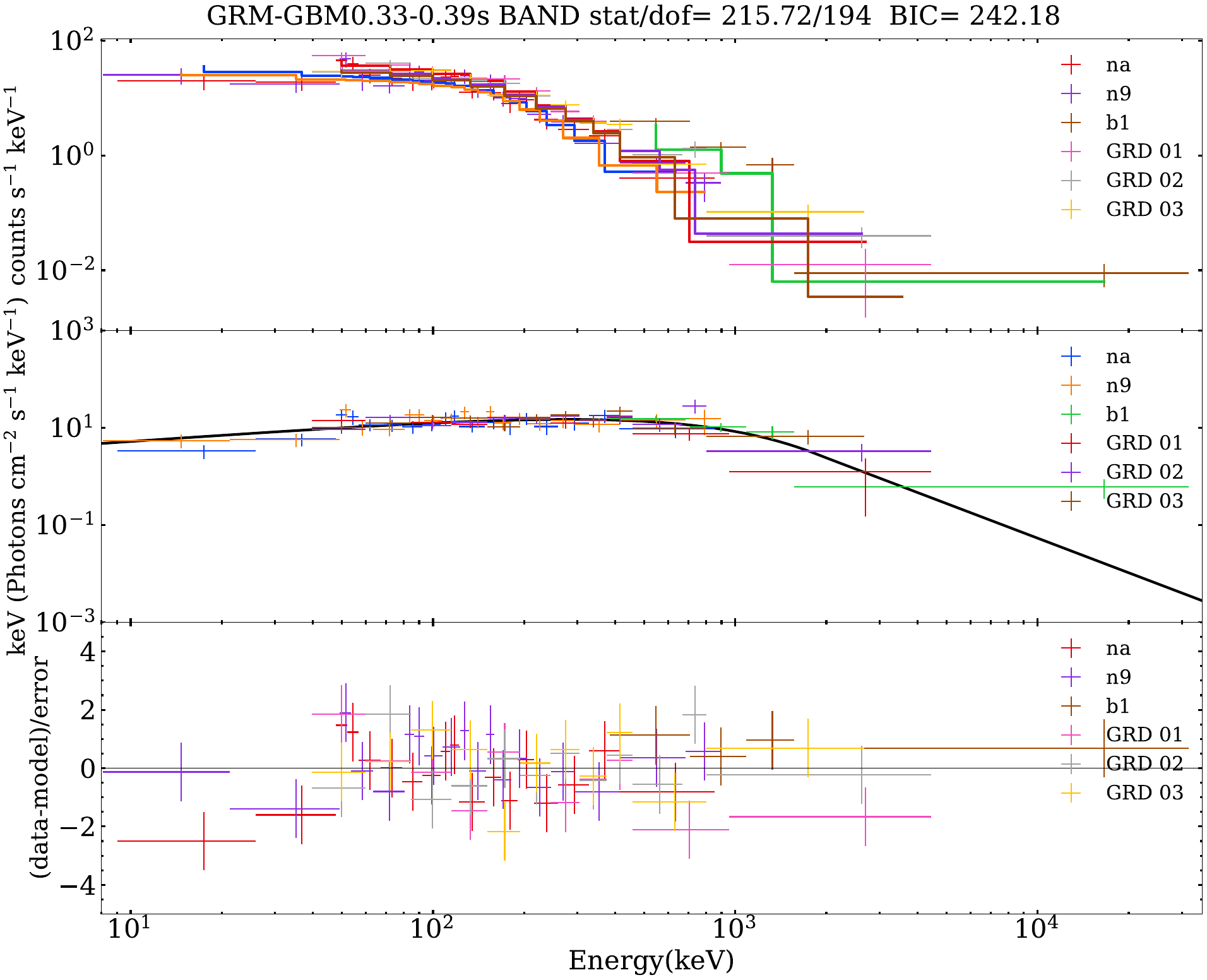}\put(0, 80){\bf a}\end{overpic} &
        \begin{overpic}[width=0.45\textwidth]{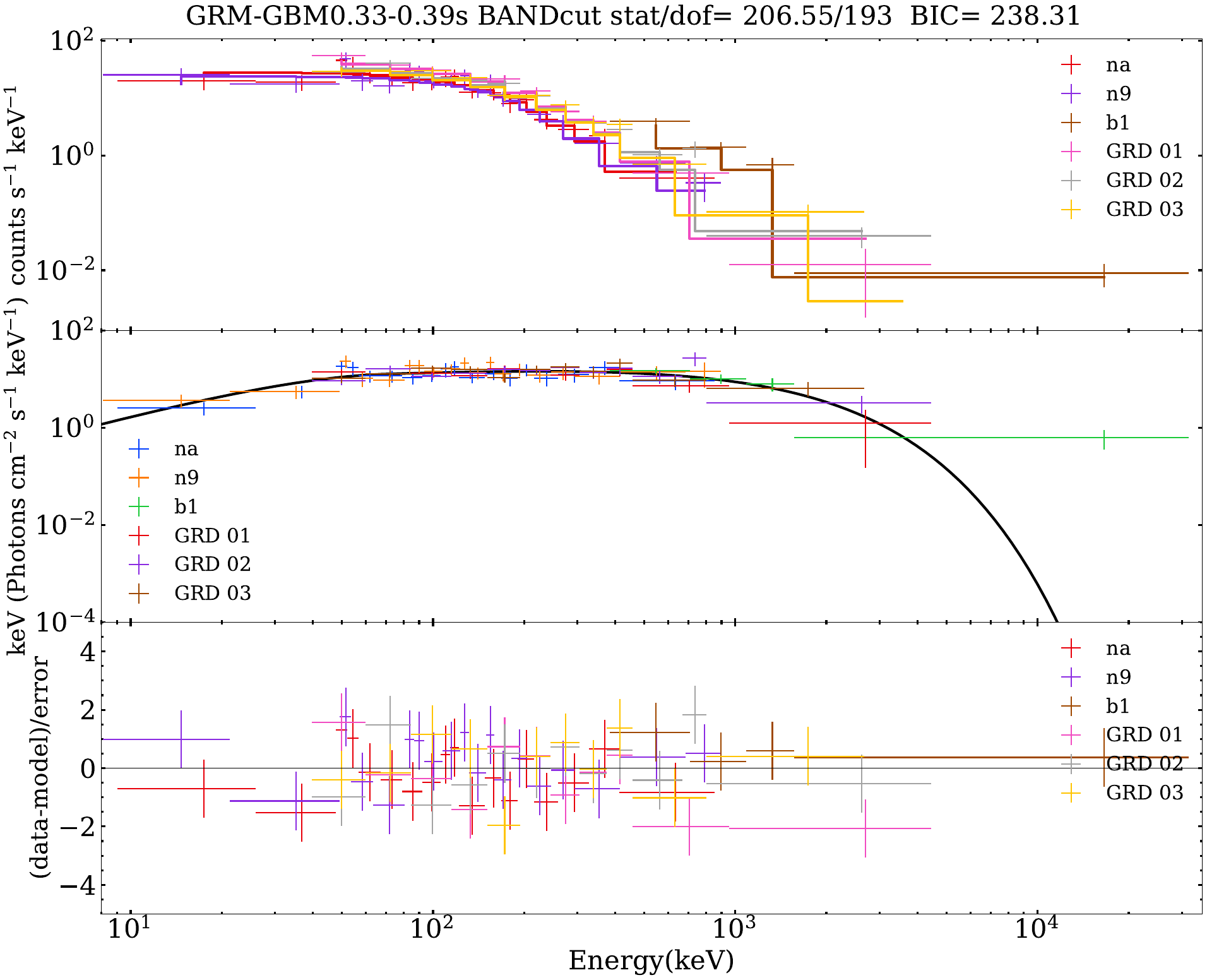}\put(0, 80){\bf b}\end{overpic} \\
\begin{overpic}[width=0.45\textwidth]{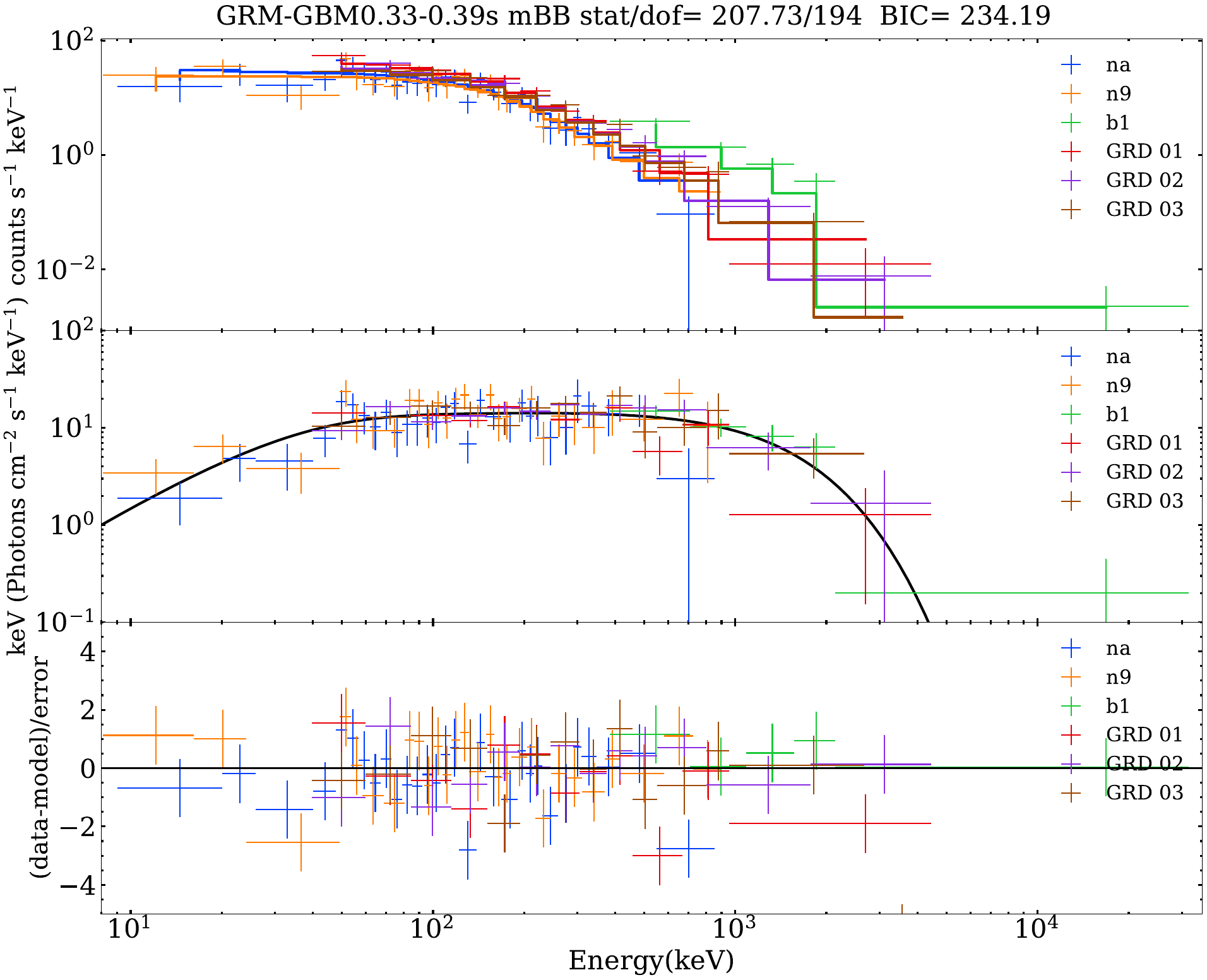}\put(0, 80){\bf c}\end{overpic} &
        \begin{overpic}[width=0.45\textwidth]{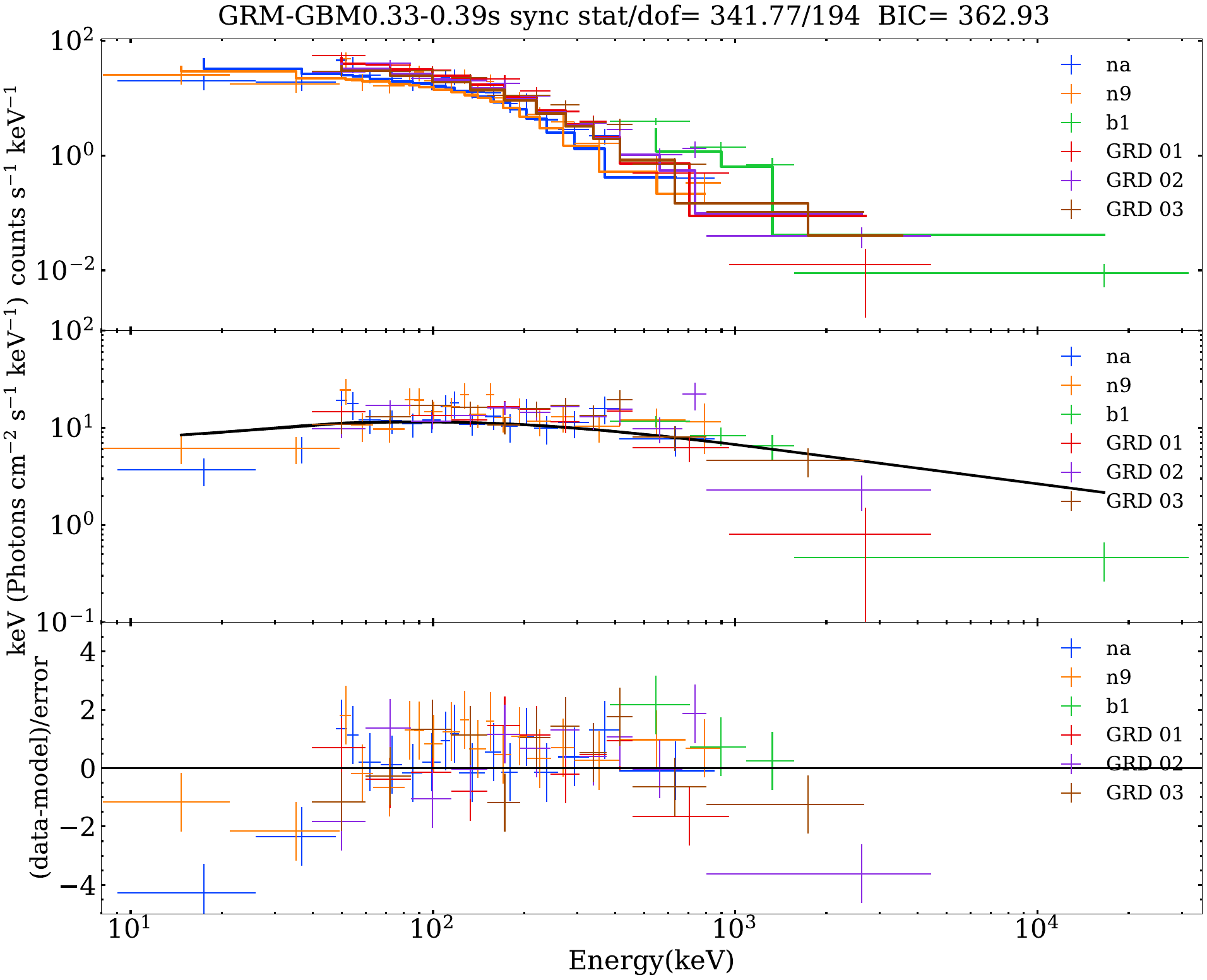}\put(0, 80){\bf d}\end{overpic} \\
\begin{overpic}[width=0.45\textwidth]{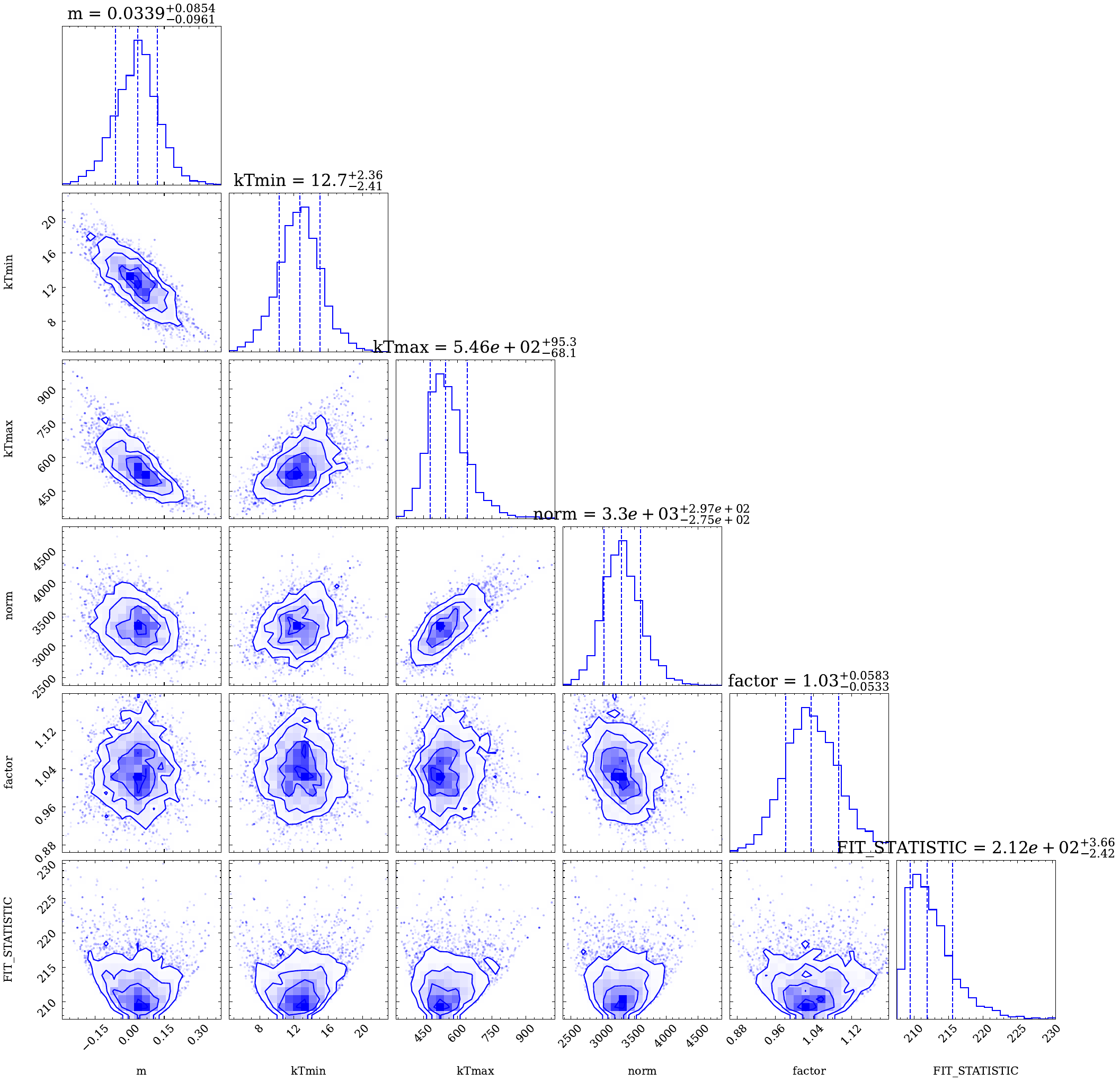}\put(0, 95){\bf e}\end{overpic} &
        \begin{overpic}[width=0.45\textwidth]{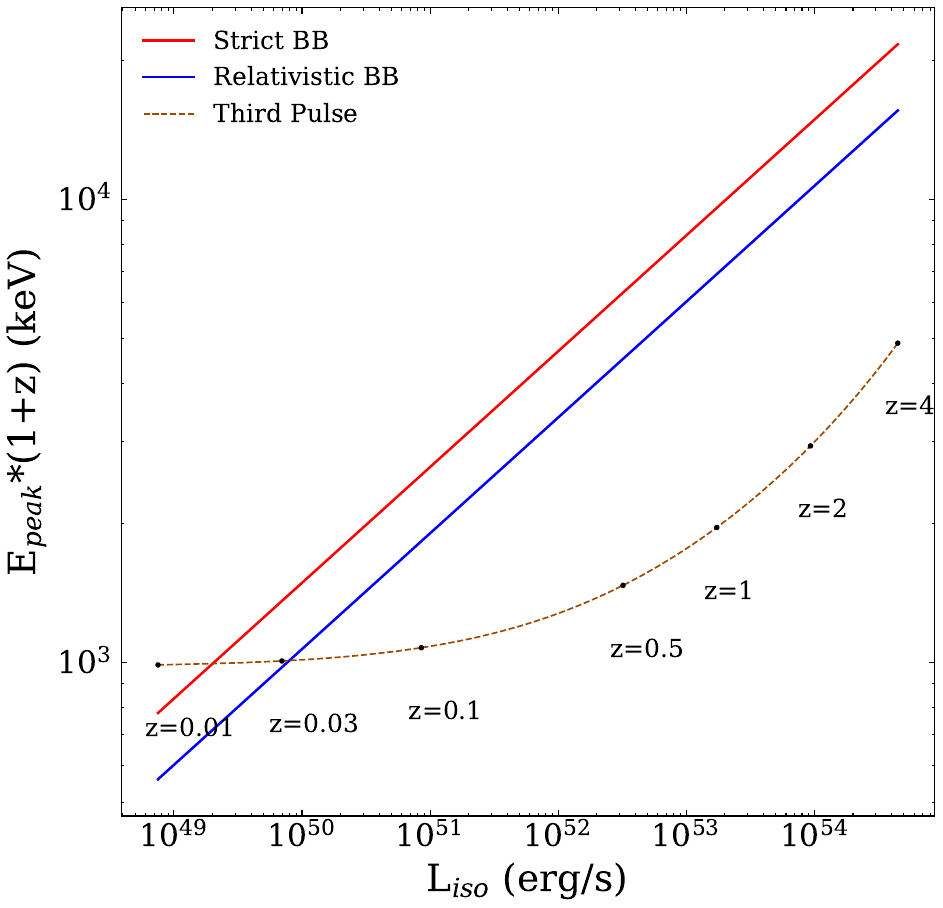}\put(0, 95){\bf f}\end{overpic} \\

\end{tabular}
\caption{\noindent\textbf{Spectral fitting results and ``death line" of photonsphere of the third pulse.}
\textbf{a-d}, The observed photon count spectrum and $F\nu$ energy spectrum of BAND model, BAND-Cut model, mBB model and synchrotron model, respectively.  \textbf{e}, The corner plot of the posterior probability distributions of the parameters of mBB model. \textbf{f}, The so-called ``photosphere death line” constraints. The red line is for a strict blackbody, and the blue line is for a relativistic multicolor blackbody outflow. The red dashed line represents the rest-frame peak energy $E_{p,z}$ and the observed isotropic $\gamma$-ray luminosity $L_{iso}$ of the third pulse as the redshift varies.}
\label{fig:third_spec}
\end{figure*}

We comprehensively use the Bayesian Information Criterion (BIC, \cite{1978BIC}) and the Akaike Information Criterion (AIC, \cite{1998AIC,1974AIC}) to perform model comparison. 
BIC$=-2\ln L+k\ln N$, while AIC$=-2\ln L+2k$, where L represents the maximum likelihood value, $k$ denotes the number of free parameters in the model, and $N$ signifies the number of data points.
For each model, BIC and AIC values are calculated. 
Among the BIC and AIC values calculated for each model, the smallest is determined as BIC$\rm_{min}$ and AIC$\rm_{min}$.
Models having $\Delta$BIC = BIC--BIC$\rm_{min}$ or $\Delta$AIC = AIC--AIC$\rm_{min}$ larger than 4 have considerably less support \citep{AIC_BIC}. The best-fit parameters for each model are listed in Table~\ref{tab:third_spec_table}, and Fig.~\ref{fig:third_spec} shows the fitting results and goodness-of-fit of each model. Combined with the energy spectrum fitting results, we found that the following facts can explain that the energy spectrum of the third pulse could be photospheric emission rather than synchrotron radiation:
\begin{enumerate}[1. ]

\item The high energy photon index of BAND model is $\sim-6$, indicating that the high-energy segment of the spectrum is actually a cutoff rather than a power law extension;
\item The BAND model and the CPL model have the same low-energy photon indices, which both exceed the ``death line", and the same $E_{peak}$. The residual of the low-energy segment in BAND model has a structure (Fig.~\ref{fig:third_spec}a) below 50 keV, which indicates a break is needed at the low-energy segment. Moreover, $\Delta$AIC of CPL model is larger than 6 and both $\Delta$AIC and $\Delta$BIC of BAND model are larger than 6 compared to the mBB model, indicating these two models are not the best;
\item With BAND-Cut model, the structure of residuals at low-energy disappears because the low-energy break is introduced (Fig.~\ref{fig:third_spec}b). Between the low-energy photon index and the high-energy cutoff, the BAND-Cut model has a middle-energy photon index $\sim-0.8$. The mBB model can also fit well, and mBB model and BAND-Cut model have the same low-energy break and high-energy cutoff (Fig.~\ref{fig:third_spec}c), 
and $\Delta$AIC and $\Delta$BIC between these two models are relatively small. However, the low-energy photon index of BAND-Cut model is $\sim+0.6$, corresponding to the $F\nu$ spectrum index is $\sim+1.6$, which is consistent with the index of a blackbody spectrum with Rayleigh-Jeans approximation in low-energy segment. Moreover, BAND-Cut is a semi-empirical phenomenological model, so it is more appropriate to describe data with the mBB model, which has physical meaning. In addition, the CPL+BB model also fits the data well, but it gives a relatively large $\Delta$BIC and $\Delta$AIC, so the mBB model should be chosen instead of CPL+BB model.
\item Since the third pulse is a well-shaped single pulse, indicating that its emission originates from a single dissipation process, thus can be directly fitted with the physical model of synchrotron radiation to determine whether it is synchrotron radiation or not. We use a parametric synchrotron model, which has been developed and packaged as a Python package by  \cite{2020NaBurgess}, to do the fit.
The cool synchrotron photon spectral of this model is in the form:

\begin{equation}
    N(E)=(E;K,B,p,\gamma _{cool},\gamma_{inj},\gamma_{max})
\label{equ:sync_Model}
\end{equation}

where $K$ is the arbitrary normalization constant of flux, $B$ is the magnetic field strength (unit: Gauss), $p$ is the electron injection index, $\gamma _{cool}$ is the energy of electron cooling during the synchrotron cooling time, $\gamma_{inj}$ is the electron injection energy and $\gamma_{max}$ is the maximum electron energy. Unit of later three parameters is kiloelectronvolts. We choose to fix $\gamma_{inj} = 10^5$ and  $\gamma_{max} = 10^8$ as the former has a strong degeneracy with $B$ and the later does not affect the spectrum shape and is usually fixed in the code. The detail description can infer to  \cite{2020NaBurgess} and  \cite{2021ApJchen}. The fitting results of the synchrotron model are poor (Fig.~\ref{fig:third_spec}d): 1) Due to the limitation of the ``death line", the $F\nu$ spectrum index close to 2 at the low-energy segment cannot be described by synchrotron model. Although the index of 2 can be given by synchronous self-absorption, it generally appears in the optical and NIR band rather than the X-ray energy band; 2) The electron index $p$ is too large to be a realistic value. The $-(p-1)/2$ in the $F\nu$ spectrum is essentially the predicted spectral index of the synchrotron radiation spectrum in the energy segment between the ``death line" at the low-energy segment and the cutoff at the high-energy segment, where a large $p$ indicates that the model directly fits the high-energy cutoff and cannot describe middle-energy photon index $\sim-0.8$. In the $F\nu$ spectrum of BAND-Cut model, the low-energy spectrum index is $\sim+1.6$, and the middle-energy spectrum index is $\sim+0.2$. In fact, neither the fast cooling model nor the slow cooling model can give the prediction that the $F\nu$ spectrum will continue to rise after the low-energy ``death line" segment, which proves that the synchrotron radiation physical model cannot decribe the observed data, not to mention it gives the largest $\Delta$BIC and $\Delta$AIC.
\end{enumerate}

In summary, we believe that the best model of the third pulse is the mBB model, which indicates that the third pulse could be interpreted as a matter-dominated photospheric emission. Further, we investigated whether the fitting results exceeded the so-called ``photosphere death line” constraints. As discussed in \cite{2012phdeathline}, the photospheric emission model has an upper limit for the peak energy of a GRB defined by:

\begin{equation}
E_{p}\le  \zeta kT_{0}\simeq 1.2{\rm MeV}\zeta  L_{52}^{1/4}R_{0,7}^{-1/2} 
\label{equ:ph_deathline}
\end{equation}

where $L_{52}$ is the luminosity in unit of $10^{52}\ erg\ s^{-1}$, $R_{0}$ is the initial fireball radius in unit of $10^{7}$ cm, and the factor $\zeta$ is taken as 3.92 for a strict blackbody, and 2.82 for a relativistic multicolor blackbody outflow. Fig.~\ref{fig:third_spec}f shows the $E_{p,z} -L_{iso}$ curve for the third pulse of different redshifts, where $R_{0}$ is taken as a typical value of $10^{7}$ cm for photospheric emission. In the redshift range from 0.03 to 0.4, the $E_{p,z}$ dose not exceed the ``photosphere death line”, which further indicates that photospheric emission cannot be ruled out.

\section{Discussion}\label{section4}

\subsection{The reason of negative lag in first pulse}

In previous studies, there were few observations of negative lags, and the observed negative lags were very small, most of them were consistent with zero within the error, and some papers even thought that there should be no significant negative lag \citep{2006lulag,2014INTEGRALlag}. Therefore, the intrinsic mechanism of negative lags has been barely considered. In GRB 240715A, due to the presence of a well-shaped pulse with significant negative lag, the intrinsic mechanism of the soft to hard evolution of energy spectrum needs to be formally investigated.

As interpreted in \cite{2018LuLag}, the temporal evolution of $E_{peak}$ can Phenomenologically explain the result of spectral lag, which is actually the shift of the peak intensity of the light curve and the peak time of the peak energy. However, such an interpretation cannot explain the negative lag of the first pulse of GRB 240715A, because the energy bands where negative lag occurs are much lower than $E_{peak}$ of this pulse (Fig.~\ref{fig:lag_of_each_pulse}a), which means that the evolution of $E_{peak}$ has no significant influence on negative lag of the first pulse. This is inconsistent with the previous work in which $E_{peak}$ evolution leading to the lag \citep{2005rydelag,2003Kocevskilag,2011penglag,2000wulag,2018LuLag}, and provides a new alternative perspective on the emergence of lag: the evolution of the low-energy spectral index $\alpha$ can also lead to the generation of lag \citep{2007HafiziLLR,2016MochkovitchLLR,2009Ghirlandalag}. Specifically for GRB 240715A, as the low-energy spectral index $\alpha$ hardens with time (Fig.~\ref{fig:spec}a), more high-energy photons arrive later, causing the observed negative lag of the first pulse. To illustrate this assumption, we perform a simple but validated simulation and find out that in addition to $E_{peak}$, the evolution of $\alpha$ alone can also lead to negative lag (Fig.~\ref{fig:lagsim}, see Appendix \ref{lag simulation} for details).

One of the possible intrinsic mechanisms of $\alpha$ evolving from soft to hard is the cooling of electrons in the decaying magnetic field \citep{2014zhangNP}. In an expanding jet such as GRBs, it is natural for magnetic field strength to decrease globally as the radius increases. In such a decaying magnetic field, the early injected electrons experience stronger cooling effect due to the stronger magnetic field, and have a wider distribution, whereas the later injected electrons have a narrower distribution due to weaker cooling effect in the weaker magnetic field, resulting in a hardening of the energy spectrum index, thus plausibly explaining the observed negative lag.



\subsection{Possible photospheric emission in the third pulse} 

In previous analysis, it can be seen that in GRB 240715A, synchrotron radiation appears before photospheric radiation, which is very difficult to understand. It is very rare to observe the transition between thermal component and non-thermal component in a single GRB. A representative event is GRB 160625B, which is the opposite case of GRB 240715A \citep{2018Zhang160625}. In GRB 160625B, the transition from photosphere to synchrotron radiation between well-separated emission episodes within a single GRB was observed. Generally speaking, a matter-dominated fireball produces the photospheric emission before the Poynting-flux-dominated outflow is launched. Because the wind escaping the photosphere travels more slowly than light, the emission from the colliding shells would lag behind the photosphere radiation. 

A similar situation to GRB 240715A occurs in GRB 170817A. The tail emission after the main pulse of GRB 170817A can be fitted by a blackbody model \citep{Goldstein0817,2018zhang0817}. It is explained as a nearly isotropic component from photospheric emission from a cocoon. However, unlike the third pulse of GRB 240715A, the tail of GRB 170817A is very weak and not a well-shaped pulse. Most importantly, the radiation of cocoon is isotropic, observed at a large viewing angle, whereas such a high $E_{peak}$ of the third pulse in GRB 240715A indicates that it is obviously generated by a relativistic jet observed at a small viewing angle and therefore is unlikely to be produced by cocoon.

From above discussion, it can be seen that it is very difficult to explain the quasi-thermal radiation in the third pulse of GRB 240715A. A plausible assumption is that different pulses corresponding to different episodes of central engine activity. It is possible that the engine is initially magnetized, but later becomes matter-dominated. After the magnetized central engine generates synchrotron radiation, it immediately becomes matter-dominated and produces a relativistic structured non-dissipative outflow with a large opening angle. For this wide outflow observed at viewing angle much smaller than the opening angle, a multicolor blackbody spectrum can be observed \citep{2013Lundmanph}. Under such an assumption, the time interval between the activities of the central engine should be very short, and the photospheric emission is produced immediately after the synchrotron radiation, so that the time interval between the observed two pulses can be very short, and even the photons of the two radiation components overlap with each other. Meanwhile, because the magnetic field is dissipated to a relatively weak level during the generation of synchrotron radiation in the first pulse, it is difficult to produce synchrotron radiation even if electrons are accelerated in the later jet. As a result, only clean photospheric emission is observed.

\section{Summary}\label{section5}

    GRB 240715A is the first short GRB detected by SVOM/GRM in-flight. 
    It has three pulses in the prompt emission, with different pulses having their own interesting properties. In this study, we comprehensively analyzed the characteristics of these pulses and found enlightening phenomena that reveal intrinsic mechanisms that are different from those in previous studies. 

    Observationally, the first and third pulse of GRB 240715A exhibit opposite evolution of spectral lag, which means the first pulse shows negative lag while the third pulse shows positive lag. The negative lag of the first pulse is relatively large, being an outlier in short GRB samples, and the lag itself accounts for about half of the pulse duration.

    Through spectral analysis, we found that the negative lag of the first pulse is caused by the evolution of spectral index, and is irrelevant to $E_{peak}$. One possible intrinsic mechanism is the electron cooling in the decaying magnetic field, which leads to the continuous hardening of the spectrum index and results in negative lag \citep{2014zhangNP}. However, to fully understand the intrinsic mechanism, more detailed calculations or the proposal of new models are still required.

    The spectral analysis also shows that the third pulse is more likely to be described by a quasi-thermal spectrum, which indicates the existence of photospheric emission. It is difficult to explain why photospheric emission appears after synchrotron radiation in a single GRB. Among many assumptions, the possible one is that after the magnetized central engine generates synchrotron radiation, it immediately becomes matter-dominated and produces a relativistic structured outflow with a large opening angle. For this wide jet observed at viewing angle much smaller than the opening angle, a multicolor blackbody spectrum can be observed. This is just a simple assumption and a more persuasive model requires more profound theoretical research. Future joint observation from ECLAIRs and GRM for more samples would better depict the low-energy spectral shape of GRBs systematically, which is also one clear asset of SVOM compared to current GRB missions \citep{SVOMprospect}
    
    Based on its outstanding performance and rapid multi-wavelength follow-up capabilities, SVOM will provide more comprehensive and precise measurements for such GRBs with novel and unique characteristics. The SVOM mission will fully explore the scientific potential of transient astrophysical phenomena, and play a crucial role in time-domain and multi-messenger astronomy. 
    
\section*{Acknowledgments}

The Space-based multi-band Variable Objects Monitor (SVOM) is a joint Chinese-French mission led by the Chinese National Space Administration (CNSA), the French Space Agency (CNES), and the Chinese Academy of Sciences (CAS). We gratefully acknowledge the unwavering support of NSSC, IAMCAS, XIOPM, NAOC, IHEP, CNES, CEA, and CNRS.

This study is supported by the National Key R$\&$D Program of China (2024YFA1611703). This work is supported by the National Natural Science Foundation of China (12273042, 12494572), the SVOM project (a mission under the Strategic Priority Program on Space Science of the Chinese Academy of Sciences), the Strategic Priority Research Program of the Chinese Academy of Sciences (Grant No. XDA30050000, XDB0550300) and the National Key R\&D Program of China (2021YFA0718500). We appreciate the public data and software of \textit{Fermi}/GBM. We appreciate the software GECAMTools-v20240514 provided by Peng Zhang.
 
\begin{table*}[htbp]
\caption{\centering{Fitting Parameters of three pulses with Triple-Gaussian Function}}
\begin{tabular*}{\hsize}{@{}@{\extracolsep{\fill}}cccccc}
\toprule
Pulse No.& Parameters & 15--50 keV& 50--200 keV & 200--500 keV&500--2000 keV \\
\hline
 & $N_{1}$$\rm(100\,cnts\,s^{-1})$ & 59.9$^{+5.4}_{-5.8}$ & 106$^{+8.4}_{-15}$& 57.8$^{+6.5}_{-2.1}$ &19.1$^{+5.1}_{-5.2}$ \\
 $P_{1}$& $t_{p1}$(ms) & 250$^{+2.4}_{-2.5}$ & 255$^{+3.2}_{-3.1}$ & 265$^{+1.5}_{-2.2}$ & 268$^{+5.0}_{-4.0}$\\
 & $s_{1}$$\rm(ms^2)$ & 756.0$^{+140}_{-149}$ & 694.0$^{+139}_{-125}$ & 344.0$^{+110}_{-318}$ &414.0$^{+229}_{-188}$ \\
 \hline
 & $N_{2}$$\rm(100\,cnts\,s^{-1})$ & 22.6$^{+3.2}_{-1.9}$ & 77.0$^{+7.0}_{-7.0}$& 38.9$^{+4.8}_{-5.2}$ & 19.7$^{+3.6}_{-4.2}$\\
 $P_{2}$& $t_{p2}$(ms) & 310$^{+6.9}_{-9.5}$ & 305$^{+4.3}_{-5.5}$ & 303$^{+1.9}_{-8.7}$&305$^{+8.9}_{-8.9}$ \\
 & $s_{2}$$\rm(ms^2)$ & 804.0$^{+143}_{-269}$ & 913.0$^{+615}_{-317}$ & 811.0$^{+135}_{-371}$ &1410$^{+508}_{-524}$ \\
 \hline
 & $N_{3}$$\rm(100\,cnts\,s^{-1})$ & 38.4$^{+4.2}_{-4.3}$ &183$^{+4.7}_{-1.9}$  & 82.3$^{+5.9}_{-7.0}$ & 15.8$^{+2.9}_{-4.2}$\\
 $P_{3}$& $t_{p3}$(ms) & 376$^{+4.0}_{-4.1}$ & 361$^{+1.1}_{-1.1}$ & 355$^{+2.0}_{-1.9}$ & 348$^{+4.2}_{-5.2}$ \\
 & $s_{3}$$\rm(ms^2)$ & 931.0$^{+51.4}_{-105}$ & 482$^{+62}_{-65}$ & 367$^{+67}_{-64}$& 428.0$^{+233}_{-161}$\\
\hline
 Bkg& $B$$\rm(100\,cnts\,s^{-1})$ & 25.6$^{+1.1}_{-1.1}$ & 14.6$^{+0.89}_{-0.86}$ & 2.66$^{+0.36}_{-0.36}$ & 1.83$^{+0.29}_{-0.25}$\\
 & $Stat/d.o.f$ & 1.33 & 1.71 & 1.46& 3.61\\
 
\botrule
\end{tabular*}
\label{tab:Pulse_fit_pars}
\end{table*}

\begin{table*}[htbp]
\caption{\centering{The spectral results of GRB 240715A}}
\begin{tabular*}{\hsize}{@{}@{\extracolsep{\fill}}ccccc}
\toprule
t$_{1}(s)$&t$_{2}(s)$ & $\alpha$ & $E_{peak}$(keV) & PGSTAT/d.o.f \\
\hline
\textbf{S-I} \\
0.19 & 0.28& -1.03$^{+0.04}_{-0.04}$ & 2949.30$^{+480.88}_{-645.05}$ & 240.44/195 \\
0.28 & 0.33 & -0.65$^{+0.05}_{-0.05}$ & 2936.46$^{+304.89}_{-341.89}$ & 238.66/195  \\
0.33 & 0.39 & -0.60$^{+0.06}_{-0.06}$ & 975.37$^{+87.38}_{-106.55}$ & 216.45/195 \\
\hline
\textbf{S-II} \\
0.19 & 0.23& -1.49$^{+0.10}_{-0.09}$ & 888.87$^{+237.03}_{-162.75}$ & 202.32/195 \\
0.23 & 0.26 & -1.12$^{+0.04}_{-0.05}$ & 3231.48$^{+560.93}_{-557.97}$ & 222.76/195  \\
0.26 & 0.29 & -0.64$^{+0.07}_{-0.07}$ & 2608.33$^{+462.37}_{-371.52}$ & 227.21/195  \\
0.29 & 0.32& -0.62$^{+0.07}_{-0.08}$ & 2419.90$^{+404.6}_{-259.69}$ & 224.59/195 \\
0.32 & 0.34 & -0.55$^{+0.07}_{-0.08}$ & 2743.66$^{+417.35}_{-318.59}$ & 203.94/195 \\
0.34 & 0.37 & --0.43$^{+0.09}_{-0.11}$ & 836.78$^{+108.52}_{-90.37}$ & 208.62/195 \\
0.37 & 0.41 & -0.27$^{+0.28}_{-0.22}$ & 107.98$^{+12.9 }_{-11.72}$ & 139.77/195  \\

\botrule
\end{tabular*}
\label{tab:spectral_of_GRB}
\end{table*}

\begin{table*}[htbp]
\caption{\centering{Fitting results of each models for the third pulse}}
\begin{tabular*}{\hsize}{@{}@{\extracolsep{\fill}}c|cccc|ccc}
\toprule
Model&& Parameters && &PGstat/dof & BIC($\Delta$BIC) & AIC($\Delta$AIC)\\
& & & & & & & \\
 \hline
&$\alpha$ &$E_{peak}$(keV)&  &&  &  & \\
CPL& & & & & & & \\
 & -0.60$^{+0.06}_{-0.06}$&975.37$^{+87.38}_{-106.55}$ & & & 216.45/195& 237.62(3.43)& 224.45(6.72)\\
 \hline
&$\alpha$ &$E_{peak}$(keV)&$\beta$ && & &  \\
BAND& & & & & & & \\
 & -0.59$^{+0.07}_{-0.06}$&999.15$^{+115.18}_{-105.43}$& -5.96$^{+2.48}_{-2.74}$& &215.72/194& 242.18(7.99)&225.72(7.99)\\
\hline
&$\alpha$  &$E_{1}$(keV)&$\beta$& $E_{2}$(keV) &  & &  \\
BAND-Cut& & & & & & & \\
 & 0.60$^{+0.19}_{-0.22}$&53.09$^{+13.65}_{-11.82}$&-0.83$^{+0.09}_{-0.07}$ & 1079.93$^{+205.58}_{-201.12}$ & 206.55/193& 238.31(4.12)&218.55(0.82)\\
\hline
&$\alpha$ &$E_{peak}$(keV)&kT(keV) & &  & &  \\
CPL+BB& & & & & & & \\
 & -0.47$^{+0.11}_{-0.11}$&1060.64$^{+158.23}_{-112.07}$ & 23.14$^{+5.83}_{-7.43}$& & 208.76/193& 240.52(6.33)& 220.76(3.03)\\
\hline
&m &$kT_{min}$(keV)&$kT_{max}$(keV) & & & &  \\
mBB& & & & & & & \\
 & 0.03$^{+0.09}_{-0.10}$&12.67$^{+2.36}_{-2.41}$ & 546.31$^{95.30}_{-68.14}$& & 207.73/194& 234.19(-)& 217.73(-)\\
\hline
&$B$(G) &$p$&$\gamma_{cool}$($10^6$keV) & &  & &  \\
Synchrotron& & & & & & & \\
 & 1445.05$^{+165.02}_{-173.81}$&5.17$^{+0.66}_{-1.28}$ &5.44$^{+0.18}_{-0.09}$ & & 341.77/194& 362.93(128.74)&351.77(134.04)\\

\bottomrule
\end{tabular*}
\label{tab:third_spec_table} 
\end{table*}

\clearpage

\bibliography{main}

\appendix
\renewcommand{\thefigure}{\thesection.\arabic{figure}} 
\renewcommand{\thetable}{\thesection.\arabic{table}}
\setcounter{figure}{0} 
\setcounter{table}{0}

\section{Pulse Lag sample} \label{lag sample}

\begin{table*}[htbp]
\caption{\centering{The lag and width of pulses of sGRBs}}
\begin{tabular*}{\hsize}{@{}@{\extracolsep{\fill}}ccc|ccc}
\toprule
 \hline
 GRB & pulse width  & lag  & GRB & FWHM  & lag  \\
 &(ms)&(ms)&&(ms)&(ms)\\
\hline
bn130504314 & $97.44\pm40.20$ & $3.00^{+8.00}_{-12.00}$ & bn150922234 & $37.24\pm16.57$ & $2.00^{+5.00}_{-6.00}$ \\
bn130504314 & $166.15\pm61.32$ & $3.00^{+5.00}_{-11.00}$ & bn150922234 & $47.10\pm40.60$ & $1.00^{+1.00}_{-5.00}$ \\
bn130912358 & $52.66\pm28.71$ & $-6.50^{+8.00}_{-11.50}$ & bn150922234 & $44.06\pm15.72$ & $0.00^{+5.00}_{-3.00}$ \\
bn130912358 & $74.47\pm43.85$ & $-1.50^{+21.00}_{-10.50}$ & bn160726065 & $37.24\pm16.57$ & $-6.00^{+6.00}_{-5.00}$ \\
bn141011282 & $13.93\pm9.08$ & $-1.00^{+0.00}_{-3.00}$ & bn160726065 & $139.32\pm74.12$ & $-4.00^{+8.00}_{-3.00}$ \\
bn141011282 & $33.30\pm14.82$ & $-4.00^{+4.00}_{-7.00}$ & bn170206453 & $235.50\pm165.75$ & $1.00^{+2.50}_{-3.50}$ \\
bn180402406 & $91.21\pm49.72$ & $4.00^{+8.00}_{-10.00}$ & bn170206453 & $372.36\pm234.40$ & $0.50^{+5.50}_{-2.50}$ \\
bn180402406 & $37.24\pm16.57$ & $1.00^{+9.00}_{-7.00}$ & bn170206453 & $139.32\pm74.12$ & $6.50^{+7.00}_{-4.00}$ \\
bn180703949 & $235.50\pm74.12$ & $15.00^{+19.00}_{-21.00}$ & bn191227723 & $33.30\pm9.08$ & $0.00^{+3.00}_{-1.50}$ \\
bn180703949 & $407.90\pm90.78$ & $13.00^{+13.00}_{-18.00}$ & bn191227723 & $37.24\pm7.41$ & $0.00^{+1.00}_{-4.00}$ \\
bn180728882 & $33.30\pm16.57$ & $9.00^{+13.00}_{-9.00}$ & bn191227723 & $44.06\pm28.71$ & $-3.00^{+3.00}_{-2.00}$ \\
bn180728882 & $166.52\pm74.12$ & $2.00^{+39.00}_{-15.00}$ & bn230418883 & $52.66\pm40.60$ & $-2.00^{+1.00}_{-3.00}$ \\
bn200219317 & $210.64\pm104.83$ & $0.00^{+18.00}_{-16.00}$ & bn230418883 & $52.66\pm28.71$ & $3.00^{+5.00}_{-6.00}$ \\
bn200219317 & $166.52\pm74.12$ & $5.00^{+7.00}_{-19.00}$ & bn230418883 & $74.47\pm37.06$ & $0.00^{+5.00}_{-4.00}$ \\
bn200605762 & $105.32\pm40.60$ & $0.00^{+10.00}_{-17.00}$ & bn240821775 & $105.32\pm52.41$ & $2.00^{+8.00}_{-2.00}$ \\
bn200605762 & $91.21\pm33.15$ & $7.00^{+16.00}_{-10.00}$ & bn240821775 & $117.75\pm43.85$ & $10.00^{+15.00}_{-8.00}$ \\
bn211124104 & $47.10\pm23.44$ & $-1.50^{+6.50}_{-3.00}$ & bn240821775 & $128.99\pm52.41$ & $9.00^{+13.00}_{-13.00}$ \\
bn211124104 & $117.75\pm74.12$ & $-3.00^{+4.00}_{-15.00}$ & bn121127914 & $123.41\pm51.96$ & $-7.00^{+24.00}_{-19.00}$ \\
bn221022612 & $33.30\pm12.84$ & $1.50^{+5.00}_{-5.00}$ & bn130628860 & $33.30\pm14.82$ & $-3.00^{+4.00}_{-4.00}$ \\
bn221022612 & $16.65\pm12.84$ & $0.50^{+4.50}_{-5.50}$ & bn130628860 & $44.06\pm28.71$ & $-1.00^{+8.00}_{-12.00}$ \\
bn230906538 & $28.84\pm14.82$ & $-2.50^{+3.00}_{-5.00}$ & bn140209313 & $372.36\pm74.12$ & $16.00^{+15.00}_{-15.00}$ \\
bn230906538 & $74.47\pm28.71$ & $-5.00^{+3.00}_{-5.00}$ & bn160822672 & $20.77\pm5.70$ & $-1.00^{+1.00}_{-2.00}$ \\
bn231205101 & $72.54\pm33.15$ & $6.00^{+4.00}_{-13.00}$ & bn170708046 & $45.26\pm21.43$ & $4.00^{+6.00}_{-5.00}$ \\
bn231205101 & $113.77\pm49.72$ & $2.00^{+7.00}_{-11.00}$ & bn171108656 & $20.84\pm7.46$ & $2.00^{+1.00}_{-3.00}$ \\
bn240810880 & $74.47\pm33.15$ & $0.50^{+5.50}_{-20.00}$ & bn190427190 & $24.60\pm12.93$ & $-2.00^{+2.00}_{-4.00}$ \\
bn240810880 & $235.50\pm90.78$ & $13.00^{+19.50}_{-24.50}$ & bn201227635 & $47.85\pm17.36$ & $1.00^{+3.00}_{-4.00}$ \\
bn130416770 & $13.93\pm7.41$ & $0.00^{+4.00}_{-4.00}$ & bn230116374 & $56.24\pm19.00$ & $-1.00^{+4.00}_{-8.00}$ \\
bn130416770 & $15.80\pm7.41$ & $1.00^{+5.00}_{-2.00}$ & bn230525449 & $17.43\pm7.38$ & $1.00^{+1.00}_{-1.00}$ \\
bn130416770 & $23.55\pm14.82$ & $3.00^{+3.00}_{-6.00}$ & bn240615744 & $39.82\pm14.46$ & $3.00^{+4.00}_{-4.00}$ \\
&&& bn240625314 & $61.43 \pm22.14$ & $17.00^{+2.00}_{-7.00}$\\ 

\bottomrule
\end{tabular*}
\label{tab:lag_width} 
\end{table*}

\section{Simulation of the lag generation} \label{lag simulation}

We take the form as the photon flux density $N(t, E)$ observed at a given time $t$ and photon energy $E$ as:

\begin{equation}
    N(E,t) = I(t)*\phi (E,t)
\label{equ:sim_spec}
\end{equation}

where $I(t)$ and $\phi(E, t)$ are the intensity and normalized photon spectrum, respectively. We first take $I(t)$ as the same Gaussian profile as first pulse in GRB 240715A:

\begin{equation}
    I(t) = A*exp\left [ - \frac{({t-\mu )}^{2}}{\sigma} \right ],
    \label{eq:Gaussian}
\end{equation}

Then we take $I(t)$ as the FRED profile \citep{Norris05}
for generalization:

\begin{equation}
    I(t) =  \frac{A*\lambda}{exp(\frac{\tau_{\rm{r}}}{t}+\frac{t}{\tau_{\rm{d}}})},
    \label{eq:Norris05}
\end{equation}

 where $\lambda = exp(2\mu)$, $\mu = \sqrt{\tau_{\rm{r}}/\tau_{\rm{d}}}$,  $\tau_{\rm{r}}$ and $\tau_{\rm{d}}$ are the rising and decaying time scales, respectively. For Gaussian case, we take $\phi(E, t)$ as the CPL model in order to be sonsistent with the observation. For FRED case, We take $\phi(E, t)$ as the BAND function \citep{Band1993ApJ} with its high-energy photon index $\beta$, fixed to -2. The $E_{peak}$ is fixed to 3000 keV for both spectrum models. For the low-energy photon index, $\alpha$, we take it as a linear evolution over time from soft to hard, just consistent with the spectral result in section \ref{section3.3}, i.e.:
 \begin{equation}
    \alpha(t) =  \alpha_{0}+k_{\alpha}(t-t_{0})  \qquad k_{\alpha} > 0  ,
    \label{eq:alpha}
\end{equation}

\begin{figure*}[ht]
\centering
\begin{tabular}{cc}
\begin{overpic}[width=0.54\textwidth]{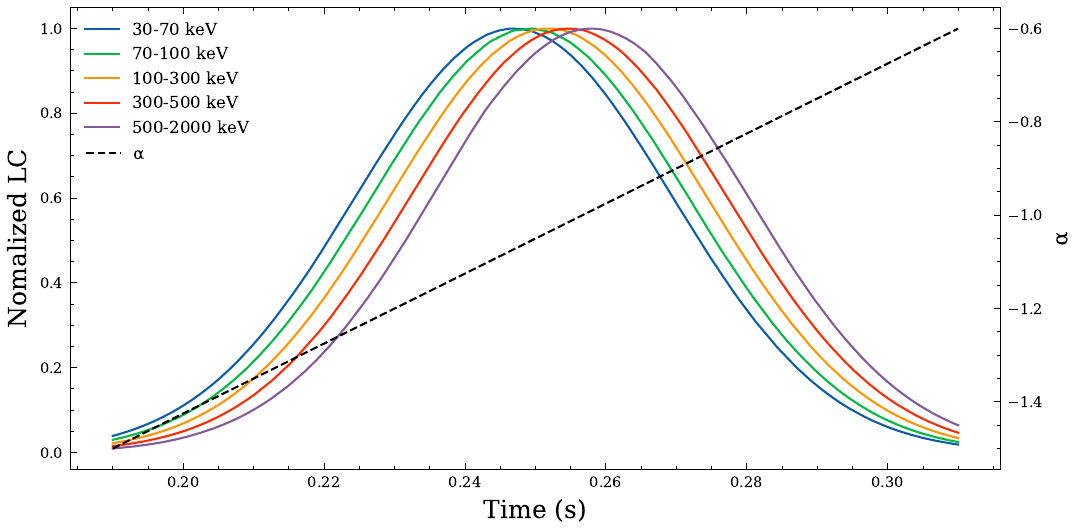}\put(0, 48){\bf a}\end{overpic} &
\begin{overpic}[width=0.36\textwidth]{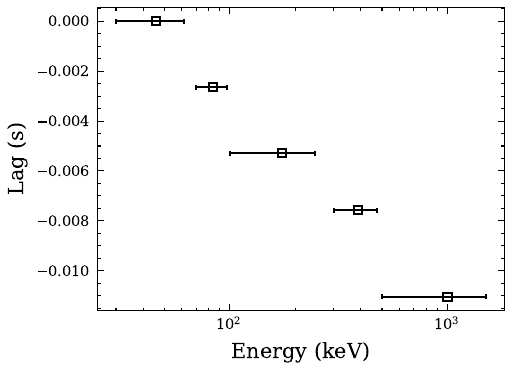}\put(0, 70){\bf b}\end{overpic} \\
\begin{overpic}[width=0.54\textwidth]{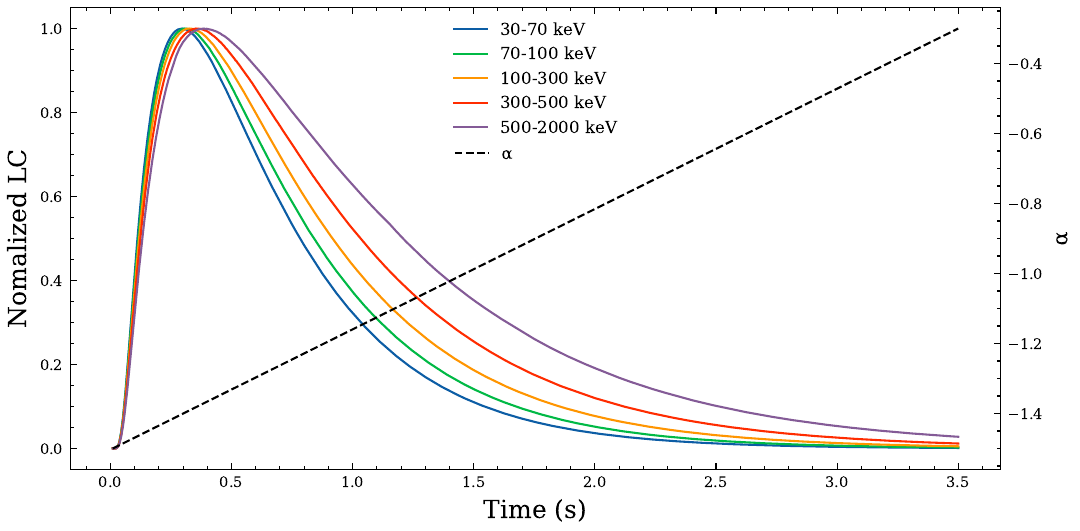}\put(0, 48){\bf c}\end{overpic} &
\begin{overpic}[width=0.36\textwidth]{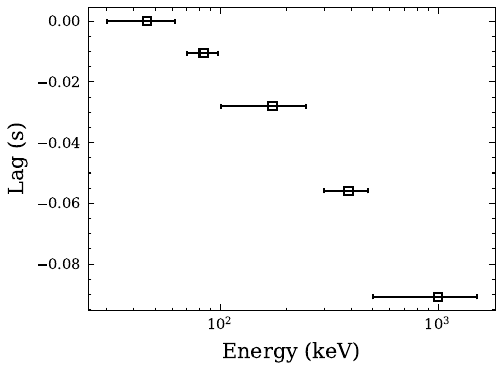}\put(0, 70){\bf d}\end{overpic} \\
\end{tabular}
\caption{\noindent\textbf{Simulation of the lag generation.}
\textbf{a} Simulated Gaussian light curves in defferent enrgy bands along with the $\alpha$ evolves from soft to hard. The parameters of $A = 60, \mu =0.25,\sigma =0.001, \alpha_{0} = -1.5, k_{\alpha} = 7.3 \,\ and\,\ t_{0} = 0.19s$ are adopted to be consistent with the obsevations. \textbf{b}, The lag versus energy of the simulated Gaussian light curves. \textbf{c}, Simulated FRED light curves in defferent energy bands along with the $\alpha$ evolves from soft to hard. The parameters of $A = 60, \tau_{\rm{r}}=0.2,\tau_{\rm{d}}=0.5, \alpha_{0} = -1.5, k_{\alpha} = 0.2 \,\ and\,\ t_{0} = 0$ are adopted.  \textbf{d}, The lag versus energy of the simulated Gaussian light curves. }
\label{fig:lagsim}
\end{figure*}

 In order to avoid the influence of $E_{peak}$ and only investigate the influence of the evolution of $\alpha$ of the spectrum, five energy bands the same as those used to calculate lag from 30 keV to 2000 keV are chosen to calculate the light curves. The simulated light curves are displayed in Fig.~\ref{fig:lagsim} for Gaussian and FRED case, respectively. In Gaussian case, the simulation results are quite consistent with the observations within the error.

\clearpage

\bibliographystyle{aasjournalv7}

\end{document}